\documentclass[twocolumn]{aastex62}
\usepackage{graphicx}

\usepackage{comment, verbatim}
\usepackage{amsmath}

\newcommand{\ha}{\ensuremath{\textrm{H}\alpha}}

\newcommand{\kms}{\ensuremath{\textrm{km s}^{-1}}}
\newcommand{\cm}{\ensuremath{ \, \mathrm{cm}}}
\newcommand{\dg}{\ifmmode{^{\circ}}\else $^{\circ}$\fi}
\newcommand{\msun}{\ensuremath{\textrm{M}_\sun}}

\newcommand{\hi}{\ion{H}{1}}

\newcommand{\lb}{\ifmmode{(\ell,b)} \else $(\ell,b)$\fi}
\newcommand{\nii}{[\ion{N}{2}]}

\newcommand{\sii}{[\ion{S}{2}]}

\newcommand{\vlsr}{\ifmmode{\rm v_{\rm{LSR}}}\else $\rm v_{\rm{LSR}}$\fi}
\newcommand{\vlmcsr}{\ifmmode{\rm v_{\rm{LMCSR}}}\else $\rm v_{\rm{LMCSR}}$\fi}
\newcommand{\av}{\ifmmode{A(V)}\else $A(V)$\fi}
\newcommand{\ebv}{\ifmmode{E(B-V)}\else $E(B-V)$\fi}
\newcommand{\iha}{\ensuremath{I_{\mathrm{H}\alpha}}}
\newcommand{\nhi}{\ensuremath{N_{\rm{H\,\sc{I}}}}}

\newcommand{\ihb}{\ifmmode{I_{\rm{H}\beta}} \else $I_{\rm H \beta}$\fi}
\newcommand{\isii}{\ifmmode{I_{\ion{\rm{S}}{2}}} \else $I_{\rm [S \textsc{ ii}]}$\fi}
\newcommand{\inii}{\ifmmode{I_{\ion{\rm{n}}{2}}} \else $I_{\rm [N \textsc{ ii}]}$\fi}
\newcommand{\ioi}{\ifmmode{I_{\ion{\rm{o}}{1}}} \else $I_{\rm [O \textsc{ i}]}$\fi}

\shorttitle{The Diffuse Ionized Gas Halo of the Large Magellanic Cloud} 
\shortauthors{Smart et al.}

\begin{document}

 \defcitealias{Smart2019}{SHB19} 
 \defcitealias{Barger2013}{BHB13} 
 \defcitealias{Ciampa2020}{CBL20} 

\title{The Diffuse Ionized Gas Halo of the Large Magellanic Cloud}

\correspondingauthor{B. M. Smart}
\email{drbsmart@uw.edu}

\author[0000-0002-5012-3549]{B. M. Smart}
\affil{Department of Astronomy, University of Washington, Seattle, WA 98195, USA}
\affil{Department of Physics, Astronomy, and Mathematics, University of Hertfordshire, Hatfield AL10 9AB, UK}
\affil{Department of Astronomy, University of Wisconsin-Madison, Madison, WI 53706, USA}

\author[0000-0002-9947-6396]{L. M. Haffner}
\affil{Department of Physical Sciences, Embry-Riddle Aeronautical University, Daytona Beach, FL 32114-3900, USA}
\affil{Department of Astronomy, University of Wisconsin-Madison, Madison, WI 53706, USA}

\author[0000-0001-5817-0932]{K. A. Barger}
\affiliation{Department of Physics \& Astronomy, Texas Christian University, Fort Worth, TX 76129, USA}

\author[0000-0002-1295-988X]{D. A. Ciampa}
\affiliation{Department of Physics \& Astronomy, Texas Christian University, Fort Worth, TX 76129, USA}

\author[0000-0001-7301-5666]{A. S. Hill}
\affil{Department of Computer Science, Math, Physics, and Statistics, University of British Columbia, Okanagan Campus, Kelowna, BC V1V 1V7, Canada}
\affil{Space Science Institute, Boulder, CO 80301, USA}
\affil{Dominion Radio Astrophysical Observatory, Herzberg Program in Astronomy and Astrophysics, National Research Council Canada, Penticton, BC, Canada}

\author[0000-0002-7955-7359]{D. Krishnarao}
\affil{NSF Astronomy and Astrophysics Postdoctoral Fellow, Johns Hopkins University, Baltimore, MD, 21218, USA}
\affil{Department of Physics, Colorado College, Colorado Springs, CO, USA}
\affil{Space Telescope Science Institute, Baltimore, MD 21218, USA}
\affil{Department of Astronomy, University of Wisconsin-Madison, Madison, WI 53706, USA}

\author{G. J. Madsen}
\affil{Institute of Astronomy, University of Cambridge, Madingley Road, Cambridge CB3 0HA, UK}

\begin{abstract}
The Large Magellanic Cloud (LMC) has an extensive \ha\ emission halo that traces an extended, warm ionized component of its interstellar medium. Using the Wisconsin \ha\ Mapper (WHAM) telescope, we present the first kinematic \ha\ survey of an extensive region around the LMC, from  $\lb = (264\fdg5,\,-45\fdg5)$ to $(295\fdg5,\,-19\fdg5)$, covering $+150\leq \vlsr \leq +390~\kms$. We find that ionized hydrogen exists throughout the galaxy and extends several degrees beyond detected neutral hydrogen emission $(\log{\left(N_{\rm H\textsc{~i}/\rm cm^{-2}}\right)\approx18.3})$ as traced by 21-cm in current surveys.  Using the column density structure of the neutral gas and stellar line-of-sight depths as a guide, we estimate \textbf{the upper limit} mass of the  ionized component of the LMC to be roughly $M_\mathrm{ionized}\approx(0.6-1.8)\times10^{9}\,\msun$, which is comparable to the total neutral atomic gas mass in the same region 
($M_\mathrm{neutral}\approx0.75-0.85\times10^{9}\,\msun$). Considering only the atomic phases, we find $M_\mathrm{ionized}/M_\mathrm{ionized+neutral}$, to be 46\%--68\% throughout the LMC and its extended halo. Additionally, we find an ionized gas cloud that extends off of the LMC at $\lb \approx (285\arcdeg, -28\arcdeg)$ into a region previously identified as the Leading Arm complex. This gas is moving at a similar line-of-sight velocity as the LMC and has $M_\mathrm{ionized}/M_\mathrm{ionized+neutral} =$ 13\%--51\%. This study, combined with previous studies of the SMC and  extended structures of the Magellanic Clouds, continues to suggest that warm, ionized gas is as massive and dynamically-important as the neutral gas in the Magellanic System.
\end{abstract}


\section{Introduction}

At distances of $d_\odot\approx50~{\rm kpc}$  \citep{Walker2012,Pietrzyski2013} and $d_\odot\approx60\,{\rm kpc}$ \citep{Hilditch2005}, the Large Magellanic Cloud (LMC) and Small Magellanic Cloud (SMC) provide us with an opportunity to thoroughly study multiple components of an external galaxy system in high detail.  Surveys reveal that the gas in this system has a complex morphology and kinematic structure. Interactions between the Magellanic Clouds (MCs) have stripped $> 10^9 M_\odot$ of gas out of the galaxies  \citep{Fox2014} and resulted in multiple gaseous structures extending out of the system, including several large regions designated as the Bridge, Leading Arm (LA), and Stream. The neutral gas component of the LMC and SMC, along with the rest of the extended Magellanic System (MSys), has been extensively  mapped in neutral gas by multiple surveys \citep{Stanimirovic2004,bruens2005,Nidever2010,Bekthi2016}. These studies provide insight into the complex history of the interactions between the LMC, SMC, and Milky Way (MW), acting as guideposts for galaxy--galaxy simulations (e.g., \citealt{Besla2007, Besla2012, Pardy2018,Williamson2021}). 
 
 \begin{figure*}[!t]
 \begin{center}
\includegraphics[width=0.6\paperwidth,clip]{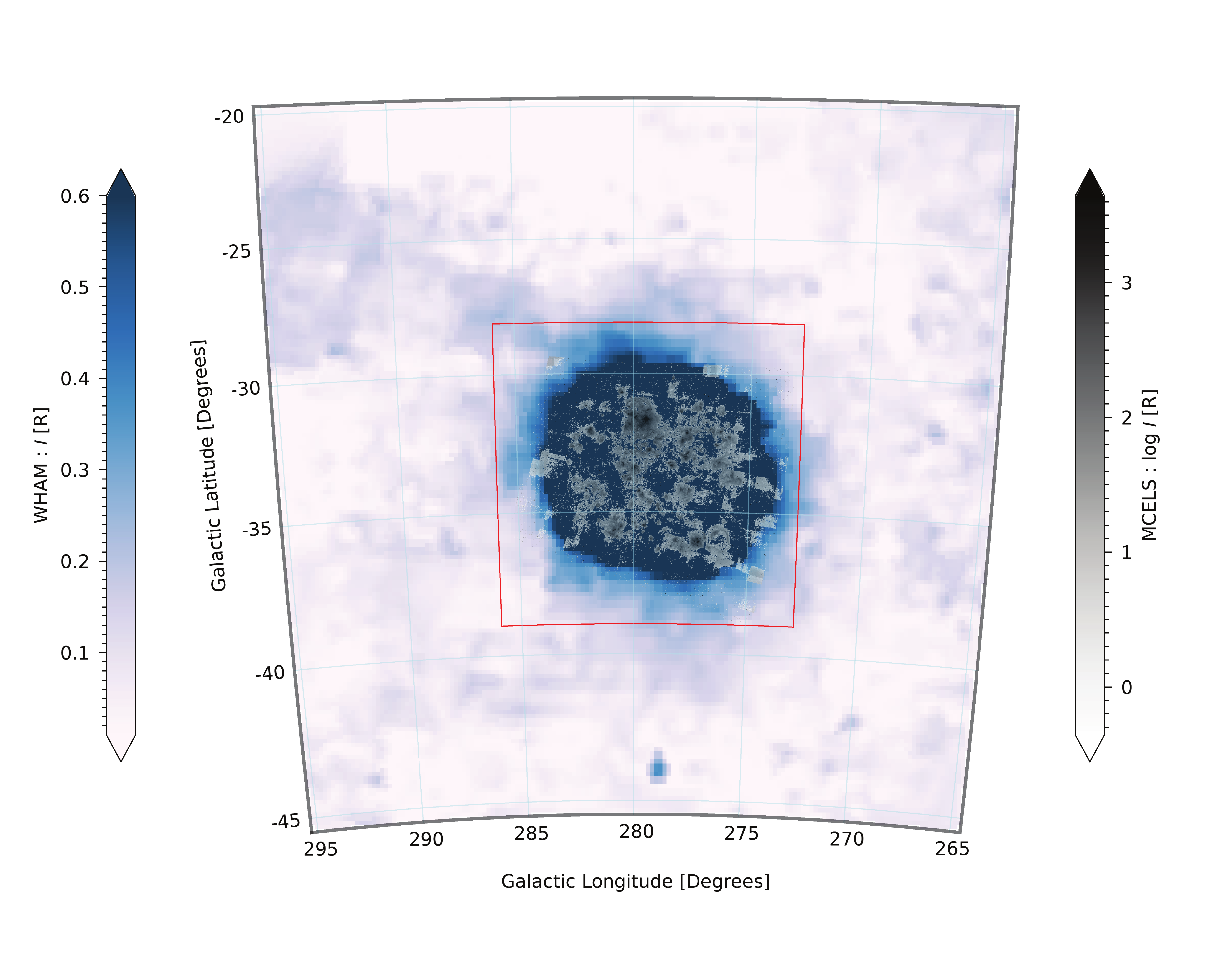}
\caption{ \ha\  emission (\emph{blue, background}) associated with the
LMC, integrated between $+150 \leq \vlsr \leq +350~\kms$
and MCELS \ha\  image
(\emph{grayscale, foreground}) from \citet{winkler2015}.
The red box denotes the extent of the MCELS survey region. Note that
the WHAM emission scaling is linear (\emph{left colorbar}) while
scaling for the MCELS image is logarithmic (\emph{right colorbar})
to highlight the bright structures within the galaxy. \label{fig:MCELS} }
\end{center}
\end{figure*}

 \begin{figure*}[!t]
 \begin{center}
\includegraphics[width=0.6\paperwidth, trim={0cm 0cm 0cm 0cm},clip]{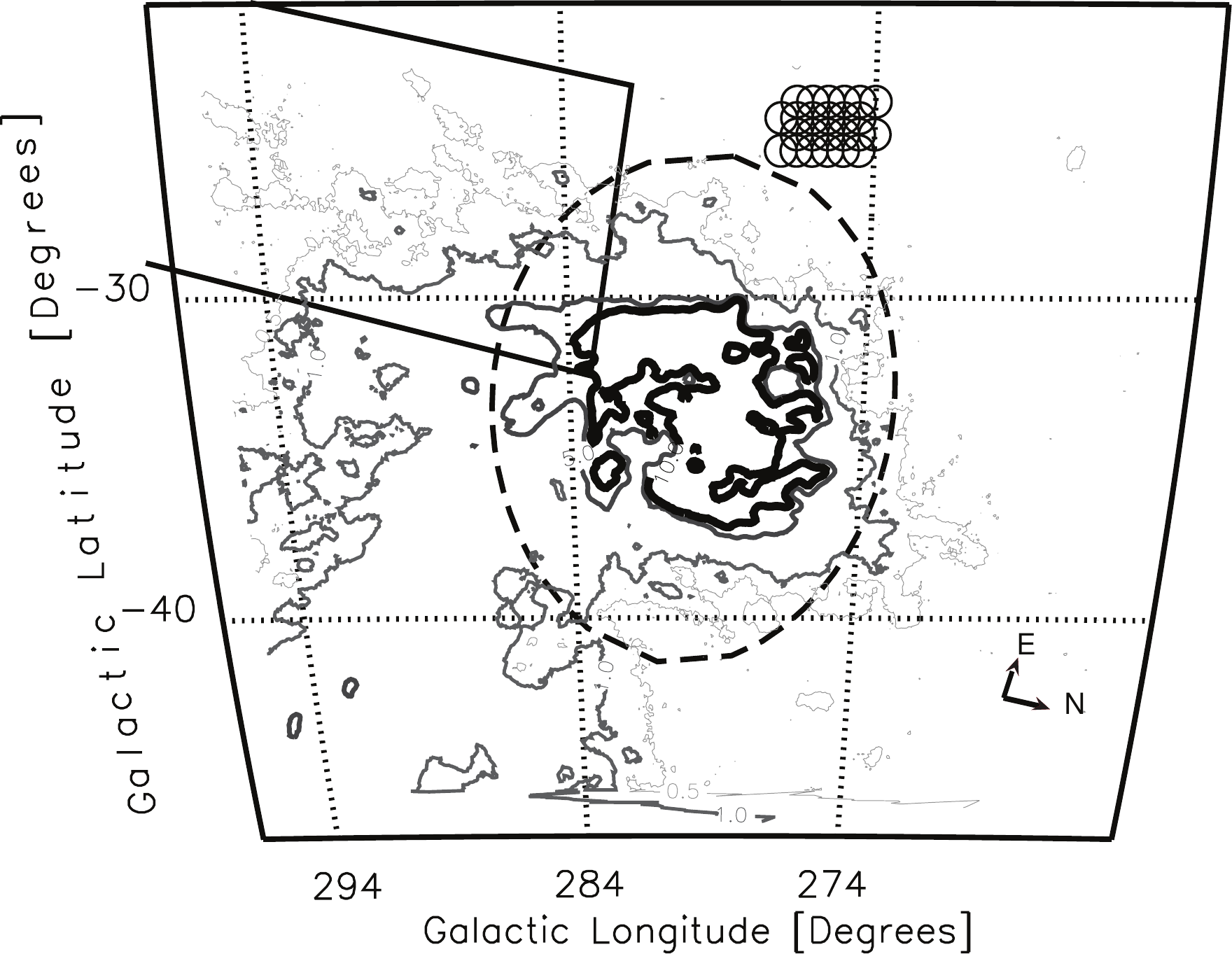}
\caption{A full map of the  LMC with each sub-region highlighted. The central LMC region is marked by the dashed ellipse. The Leading Arm (LA1.1) region at the top left is marked in solid black lines. The contours mark the \hi\ column density at $N_{\rm H\textsc{~i}}=0.5$, 1, 5 and $10\times10^{20}\,\mathrm{cm}^{-2}$. The grid of $1\arcdeg$ circles at the top left of the map represent the $0\fdg5$ Nyquist sampling used for WHAM survey observations. \label{fig:Skematic}}
\end{center}
\end{figure*} 
The distribution and kinematics of ionized gas may be a crucial input to models of the formation and history of the MCs. Models of low-mass star-forming galaxies find SMC mass galaxies may have a warm-to-cold (warm: $ 10^{4} < \mathrm{T} <10^{5}\,\rm K \mathrm{, } \, \mathrm{ cold: T} < 2000\,\rm K$) gas mass ratio of 0.7, higher than MW-mass galaxies with a ratio of 0.5 \citep{Hopkins2012}. For the LMC and SMC system, we might then expect ionized gas to form a significant fraction of the total gas mass. Studies of the dynamics and history of the MCs include observational constraints on the neutral gas \citep{Besla2007, Besla2012,Williamson2021}, but until recently had little guidance for the ionized gas content. An accurate census of the total gas content and its phases is important for improving such models.

The Magellanic Cloud Emission-Line Survey (MCELS) team imaged the bright ionized gas regions of the LMC and SMC in detail \citep{winkler2015}, which included a thorough census of the dense ionized structures and supernova remnants. This survey provides moderate spatial resolution ($\theta \lesssim 5\arcsec$) with 1-sigma \ha\ sensitivity of roughly $I_{\rm H\alpha,\,1\sigma}\approx 1.3$ R\footnote{1 Rayleigh = $\frac{10^{6}}{4\pi}$ photons s$^{-1}$ cm$^{-2}$ sr$^{-1}$} \citep{Pellegrini2012}, which can trace warm ($T_e = 10^4$ K) ionized regions with emission measures (EM) of $\gtrsim 4$ cm$^{-6}$ pc. 

While the \citet{winkler2015} study provides high resolution images of the bright \ha\ sources, a survey with higher sensitivity is needed to probe the faint warm ionized medium (WIM) emission. To gain a clearer picture of the diffuse ionized gas, the MSys has been the focus of several spectroscopically-resolved Wisconsin \ha\ Mapper (WHAM) surveys. Although WHAM observations have much lower spatial resolution of $\theta=1\arcdeg$ (roughly $1~\rm kpc$ at $D=50~\rm kpc$), the increased throughput combined with a Fabry-P{\'e}rot spectrometer provides higher sensitivity to very diffuse gas with $I_{\rm H\alpha,\,1\sigma}\approx0.03\,{\rm R}$. \textbf{Additionally, the 15 \kms\, velocity resolution of the spectra allows for identification of WIM emission. In the WIM, thermal broadening of the
\ha line in warm gas ($10^{4}$ K) combined with
non-thermal motions results in line widths $>$ 25 \kms} (see \citealt{Haffner2003}). \citet[hereafter B13]{Barger2013} surveyed the diffuse ionized gas of the Magellanic Bridge using the WHAM telescope. Their study highlighted the extent to which WHAM could detect and map the diffuse ionized gas within the MSys, and was followed with a survey of the extended ionized gas around the SMC \citep[hereafter SHB19]{Smart2019} and a survey of the LMC's galactic outflow \citep[hereafter CBL20]{Ciampa2020}. \citetalias{Barger2013}, \citetalias{Smart2019}, and \citetalias{Ciampa2020} all found diffuse ionized gas extending several degrees beyond detected \hi\ emission $(\log{\left(N_{\rm H\textsc{~i}/\rm cm^{-2}}\right)\approx18.3})$, as traced by the 21-cm line. These studies also show that a significant fraction of the atomic hydrogen is ionized. \citetalias{Barger2013} found that 36--52\% of the atomic gas in the Bridge is ionized, and \citetalias{Smart2019} found a similar ionization fraction of 42--47\% for the SMC. Both surveys compare the kinematics of the ionized and neutral components. The two phases trace each other closely in brighter regions, but \ha\ appears more kinematically decoupled in diffuse regions with velocity offsets and additional components compared to \hi. Additionally, \citet{Barger2017} and \citet{AntwiDanso2020} have detected \ha\ emission in the Magellanic Stream and Leading Arm through targeted WHAM observations. However, these complexes have yet to be thoroughly mapped.  
    
Studies of the ionized gas associated with the MSys have not been limited to the main galaxy structures. Several studies have explored the ionized gas in the circumgalactic medium of the LMC (e.g., \citealt{Wakker1998}, \citealt{Howk2002}, \citealt{Lehner2007}, \citealt{Pathak2011}, \citealt{Barger2016}, \citetalias{Ciampa2020}). On the near-side of the LMC, \citet{Howk2002} and \citet{Lehner2007} found that most of this gas is blue-shifted in their UV absorption-line study, indicating that it is flowing out of the LMC and toward the Milky Way. Similarly, \citet{Barger2016} detected a complementary outflow on the far-side of the LMC, confirming that this galaxy has a large-scale galactic wind. A correlation of this wide-spread outflowing material with young stellar activity in the LMC provides evidence that recent star formation is driving this wind (\citealt{Howk2002}, \citealt{Lehner2007}, \citealt{Barger2016}, and \citetalias{Ciampa2020}), which could also be feeding a hot LMC halo (see \citealt{Wakker1998} and \citealt{Howk2002}). \citetalias{Ciampa2020} mapped the near-side LMC outflow in \ha\ emission with WHAM and found that it has an ionized mass range of $7.4\lesssim \log{\left(M_\mathrm{ionized}/M_\odot\right)}\lesssim7.6$ on the near side of the LMC, or $\log{\left(M_\mathrm{total}/M_\odot\right)}\approx7.9$, assuming that the wind is symmetric with an ionization fraction of 75\% (see \citealt{Lehner2007} and \citealt{Barger2016}).

Using absorption-line observations toward 69~UV bright QSOs, \citet{Fox2014} found that the circumgalactic gas of the Magellanic Clouds covers 11,000 square degree across the sky and is predominantly ionized in most directions. In this extended MSys, the LMC might be a major source of gas for the Magellanic Stream as well as a potential source of material in the LA. The kinematics of the neutral gas in the LMC and the LA suggest that the LA material originated from the LMC \citep{Putman1998, Nidever2008, Venzmer2012}. Recent chemical abundance studies directed towards the LA have called this hypothesis into question \citep{Fox2018, Richter2018}, while tidal and ram-pressure models suggest an SMC origin \citep{Diaz2012, Besla2012, Yang2014}. \citet{Lucchini2020} explored the Magellanic Corona and its effects on the MSys. By including an ionized halo of gas surrounding the LMC, they predict the neutral gas structures of both the Stream and the LA. The results from these prior studies indicate more observations are needed. Observing the structure of the diffuse ionized and neutral gas can help constrain the potential origin of ionized gas in the extended features.

In this study, we present an H$\alpha$ emission map that traces the ionized gas associated with the LMC and its surrounding area, as well as its relation to the extended features of the MSys by surveying the the LMC in \ha\ emission with the WHAM telescope. We describe our observations in Section~\ref{sub:ObservationsLMC} and detail our data reduction process in Section~\ref{sub:DataReductionLMC}. In Section~\ref{sub:IntensityLMC}, we present our non-extinction corrected maps of the extended LMC system in \ha. In Section~\ref{section:Mass}, we calculate the total mass of the ionized gas considering multiple approaches for constraining the gas density and geometry. We summarize our results in Section~\ref{sub:DiscussionLMC} and present our conclusions in Section~\ref{sub:SummaryLMC}.

\section{Observations \label{sub:ObservationsLMC}}

\begin{table*}[!t]
\caption{Extinction\label{tab:Extinction} }
\begin{center}
\begin{tabular}{ccccccccc}
\hline 
\hline 
{Region} & \multicolumn{4}{c}{Foreground Extinction} &\multicolumn{4}{c}{Internal Extinction} \tabularnewline
\cline{1-9} 
 & log $\left\langle \nhi \right\rangle $ & $A(\ha)$& $\%{}_{corr}$\tablenotemark{a}  &  & log $\left\langle \nhi \right\rangle $ & $A(\ha)$ & $\%{}_{corr}$ &\tabularnewline
 \cline{1-9} 
LMC & 20.7 & 0.27 & 10\% & & 21.2 & 0.43 & 16\%\tabularnewline
Leading Arm 1.1 & 21.0 & 0.46, 0.48 & 17\%, 18\% & & \nodata & \nodata & \nodata \tabularnewline

\hline 
\end{tabular}
\end{center}

\tablenotetext{a}{We list two values that represent the effective extinction correction applied for mass calculations in Sections~\ref{sub:neno} \& \ref{sub:cylinder} and for those in Section~\ref{sub:ellipse}. These methods use different processes for determining $\left\langle \nhi \right\rangle$, and the average value from each method is displayed. Only foreground extinction for the LA 1.1 region had any appreciable difference in extinction correction when using the different methods.}

\end{table*}

The WHAM spectrometer has been optimized to detect diffuse sources of faint optical emission lines within and around the Milky Way down to $I_{\rm H\alpha}\approx25~{\rm mR}$ (EM $\approx7\times10^{-2}$ cm$^{-6}$ pc). For extended, continuous structures we can trace emission below $I_{\rm H\alpha}\lesssim10~{\rm mR}$ \citep{Barger2012}. At this sensitivity, \citetalias{Smart2019} and \citetalias{Barger2013} detected diffuse ionized gas that surrounds the SMC and lies within the Magellanic Bridge. In this study, we use this same facility to trace the ionized hydrogen both in and around the LMC.

WHAM consists of a dual-etalon Fabry-Perot spectrometer that produces a 200~\kms\ wide spectrum with 12~\kms\ velocity resolution from light spatially integrated over a 1\dg\ beam \citep{Haffner2003}. Typical \ha\ line widths from diffuse ionized gas have a FWHM$=20~\kms$, well-matched to WHAM's spectral resolution of $\Delta\rm \mathrm{v}=12~\kms$. A 30-second exposure can achieve a signal-to-noise of 20 for a $I_{\rm H\alpha}\approx0.5~{\rm R}$ line with a width of FWHM $=20~\kms$. WHAM is currently located at Cerro Tololo Inter-American Observatory in Chile.

\begin{figure}[!t]
\includegraphics[width=0.9\columnwidth,clip]{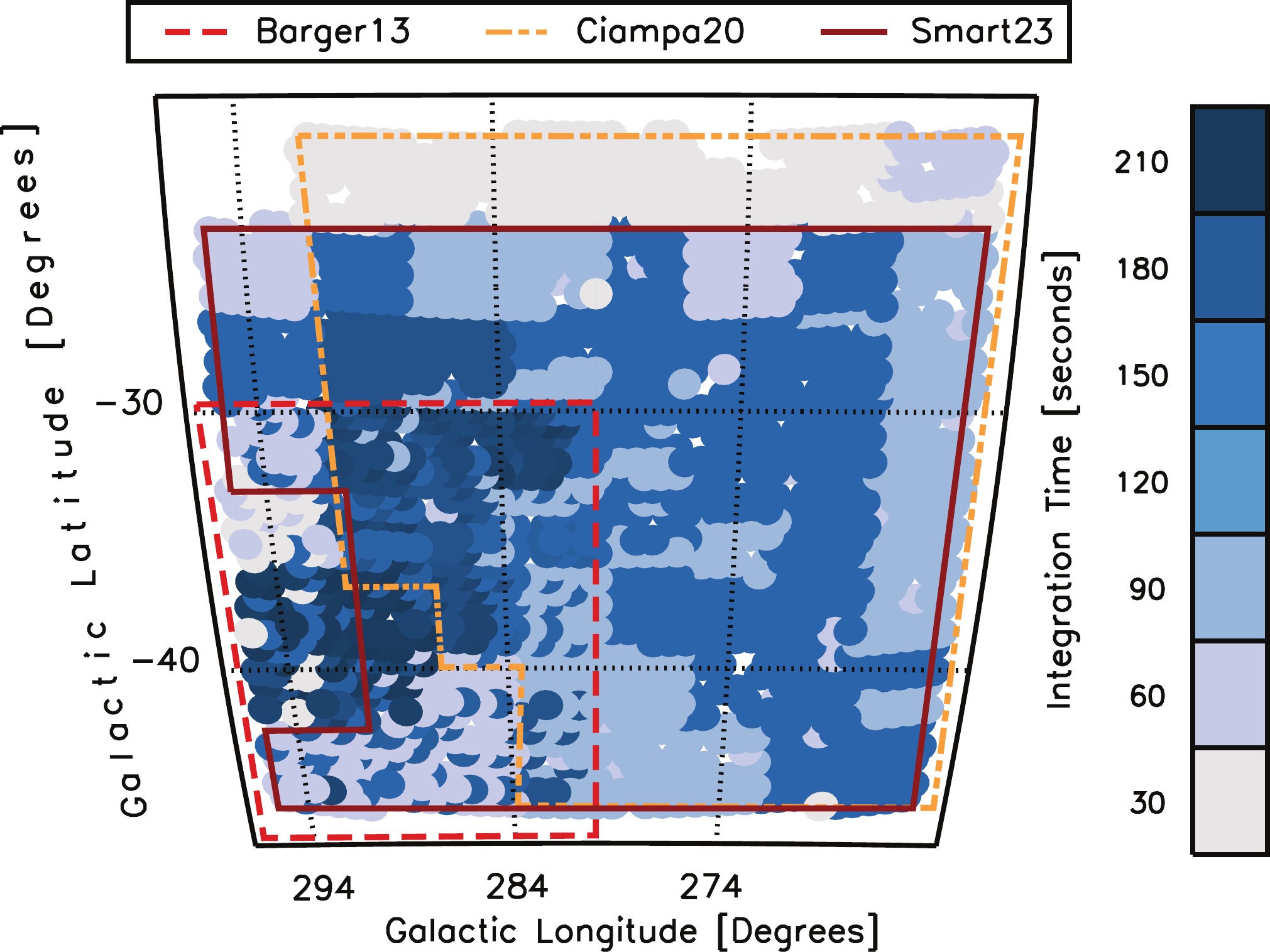}
\centering{}
\caption{Total integration time for each sightline. The solid dark red line traces the region observed in this paper. The red dashed line highlights the region observed in \citetalias{Barger2013}. The yellow dashed line indicates the region observed by \citetalias{Ciampa2020}. The final combined survey of the LMC spans a velocity range of $+150\lesssim\vlsr\lesssim+390~\kms$. \label{fig:time}}
\end{figure}

We used the same observing strategy that \citetalias{Smart2019} used for the SMC, which we summarize here. The LMC observations are grouped into ``blocks" of  30--50 Nyquist-sampled pointings with a spacing of 0.5\arcdeg. In Figure~\ref{fig:Skematic}, we illustrate the angular size and the distribution of individual pointings that comprise a typical block. To cover the extended region around the LMC, we observed 62 blocks. All but 4 of these blocks were obtained with an individual pointing exposure time of 60 seconds. The remaining 4 blocks were observed with 30-second exposures (see Figure~\ref{fig:time}). Because our sensitivity to diffuse emission is limited by contamination from very faint atmospheric lines, we kept exposure times short to track variations in the terrestrial lines throughout a block's observation. Factoring in the overhead moving to each position on the sky, a typical block with about 37 pointings and 60-second exposures had a total duration of $\sim 40$ minutes. 

Observations of the LMC were taken between January 24th and February 2nd, 2014, with 9~blocks observed or re-observed on November 23rd, 2019. This dataset focused on the local standard of rest (LSR) velocity range $+150\lesssim\vlsr\lesssim+390~\kms$ over the $264\fdg5\le l \le 295\fdg5$ and $-45\fdg5 \le b \le -22\fdg8$ region of the sky (see Figure~\ref{fig:Skematic}). 
 
We combined these observations with existing WHAM \ha\ observations of the circumgalactic medium of the MCs. Observations from \citetalias{Ciampa2020} of the LMC's near-side outflow are in the direction of the LMC but span a lower LSR velocity range, $+50\lesssim\vlsr\lesssim+250~\kms$. Existing observations of the Magellanic Bridge (\citetalias{Barger2013}) overlap with the high-longitude edge of the LMC and span a velocity range of $0\lesssim\vlsr\lesssim+315~\kms$. Data from \citetalias{Ciampa2020} between $+150\lesssim\vlsr\lesssim+250~\kms$ and from from \citetalias{Barger2013} between $+150\lesssim\vlsr\lesssim+315~\kms$ is averaged with our new LMC observations to increase signal-to-noise where they overlap. The observing strategies used in all of the combined datasets are similar (see the \citetalias{Barger2013} and \citetalias{Ciampa2020} studies for details). In Figure~\ref{fig:time}, we present the total integration time of each block used in this LMC survey after combining  new and existing observations. Outlines in Figure~\ref{fig:time} delineate the region of each of the H$\alpha$ surveys we combined for this study.

\section{Data Reduction\label{sub:DataReductionLMC}}

\subsection{WHAM Pipeline}

All observations are processed using the WHAM pipeline described in \citet{Haffner2003}.
The spectra are pre-processed in the pipeline to remove cosmic rays before combining. The Fabry-Perot spectrometer produces circular interference patterns which are
summed in annuli to produce a linear spectrum that
is a function of velocity. The pipeline then bins the spectra
in v\textsubscript{bin}= 2 \kms\ intervals. Once binned, the spectra are
normalized to exposure time and then scaled according to the airmass of the observation. The pipeline then uses an intensity
correction factor to account for sensitivity degradation
of the WHAM instrumentation that occurs over time. The standardization of the spectra allows us to directly average
observations across multiple nights and varying exposure times.

 \begin{figure*}[!t]
 \begin{center}
\includegraphics[width=0.7\paperwidth,clip]{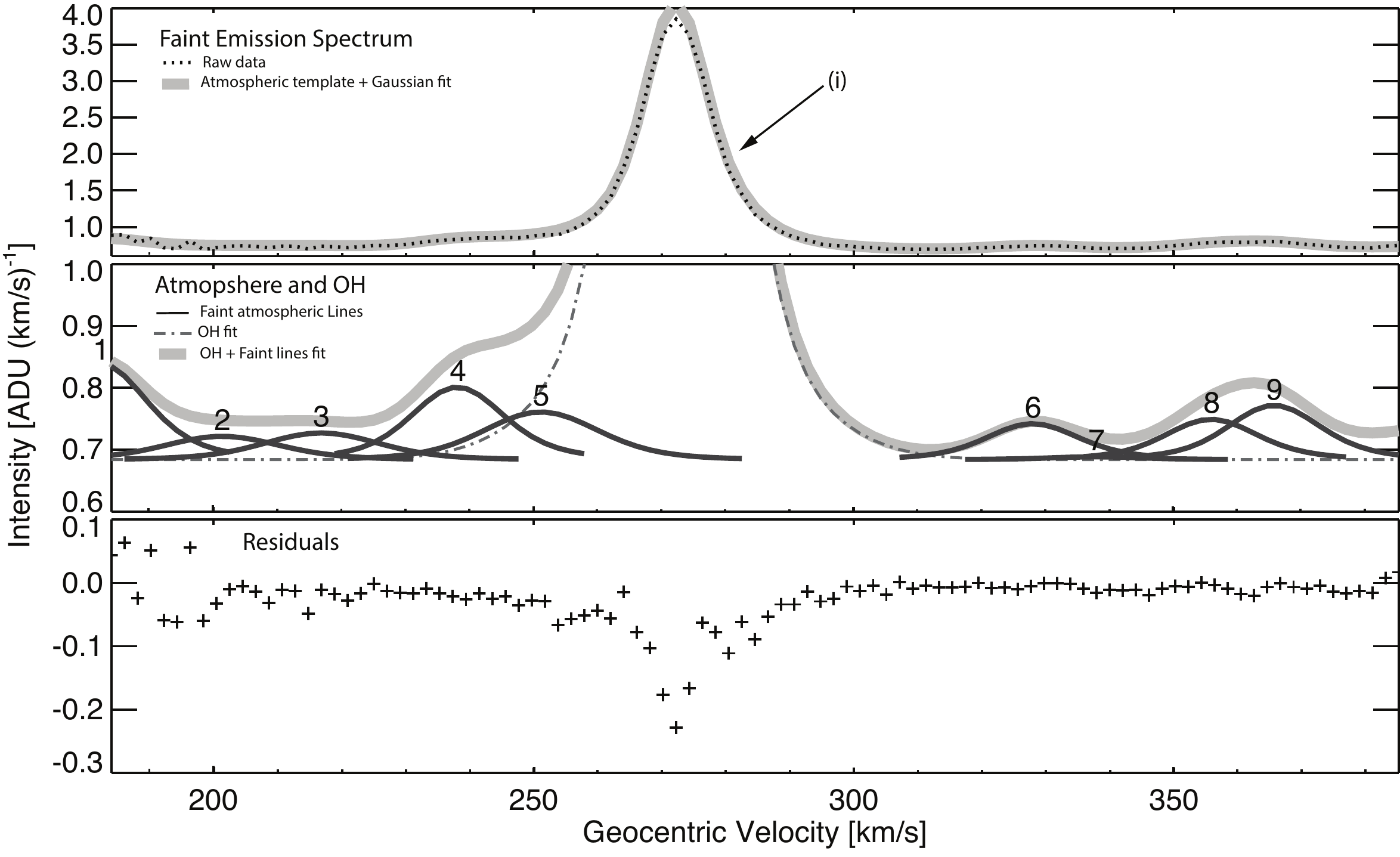}
\caption{Averaged emission template toward $(l,b) = (272\fdg5,\,-28\fdg0)$. The top and middle panels show the average spectra as a dotted line, while the light gray line outlines the atmospheric template fit. Dark gray Gaussian profiles in the middle panel trace individual atmospheric emission lines modeled by the template. The bright OH line at $\rm v_{geo}=+272.44~\kms$ is labeled \emph{(i)} in the top panel and represented as the dash-dot line in the middle panel. Residuals in the bottom panel at velocities around this bright line are due to a slight mismatch between the instrument profile and the averaged spectrum. Properties of the OH line and the atmospheric emission lines are listed in Table \ref{tab:AtmoLineTable}.
\label{fig:Atmo}}
\end{center}
\end{figure*}

\subsection{Atmospheric Line Subtraction\label{sub:Faint-Atmospheric-Line-Section}}

We follow the same methods for atmospheric line subtraction as outlined in \citetalias{Smart2019} and previous WHAM studies \citep{Haffner2003, Hill2009, Barger2012, Barger2013, Ciampa2020}. However, we observed \ha\ emission at  higher positive geocentric velocities ($v_{geo}$) than these previous WHAM surveys. To characterize the atmospheric lines at these higher velocities, we observed two \ha-faint positions on the sky located at \lb\ = $(89\fdg0, -71\fdg0)$ and \lb\ = $(272\fdg5, -28\fdg0)$. We combined 24 observations of these two sightlines taken on September 3rd, 2016 to create the atmospheric template in Figure~\ref{fig:Atmo}. The faint atmospheric emission lines are well-fit by the Gaussian components listed in Table~\ref{tab:AtmoLineTable}  and have a strength that scales with airmass.

In addition to removing the faint atmospheric lines listed in Table~\ref{tab:AtmoLineTable} with an atmospheric template, we separately removed a bright OH atmospheric line that lies at $\rm v_{geo}=+272.44~\kms$ (as seen in the top panel of Figure~\ref{fig:Atmo}) as its strength scales with the flux of sunlight on the Earth's upper atmosphere in addition to the airmass. The strength of the OH line compared to the \ha\ emission allows us to directly fit and subtract the line from our observations. 

Unfortunately, some of the faint \ha\ emission associated with the LMC can be inadvertently removed when the OH line is subtracted. The affected region  spans roughly $20\,\kms$ in width (see Figure \ref{fig:LASpectra}), wider than the unresolved core of the OH line ($\approx 12\,\kms$) due to non-Gaussian, broader wings in the instrument profile. Some of the brighter and/or broad astronomical emission can be disentangled from the OH line, but the full scope of missing emission can not be estimated well in this work using \ha\ alone. Future observations of associated emission lines at other wavelengths (e.g., \sii\ and \nii) will help reconstruct \ha\ component structure, when necessary.

Unlike many previous WHAM studies, we used a different \ha\ filter centered at longer wavelengths to capture the most positive Magellanic velocities with high transmission. However, this shift also introduces an unfortunate out-of-tune Fabry-P\'{e}rot ``ghost'' from the OH doublet at 6577.183\AA/6577.386\AA. Although these lines are at much higher velocity ($\vlsr\approx +662\,\kms$) than our target range, the combination of the filter and etalon pressure tune introduces a broad emission feature that spans roughly $+120\lesssim\vlsr \lesssim +180~\kms$ (seen as the rise in the left edge of Figure~\ref{fig:Atmo}). To avoid this feature, any sightlines that have data using only this filter are integrated over velocities greater than $\vlsr=+180~\kms$ for measurements in this work. However, sightlines combined with observations from \citetalias{Barger2013} and \citetalias{Ciampa2020} utilize the standard WHAM H$\alpha$ filter \citep{Haffner2003}, avoid this contamination. These spectroscopic observations can be fully integrated down to $\vlsr=+150~\kms$, our preferred low-velocity limit for this study. 

Finally, we use the fitting algorithm described in \citetalias{Ciampa2020} to identify systematic residuals present at specific geocentric velocities after removing the atmospheric lines, as described above. Sightlines with little to no Galactic and Magellanic emission are then used to correct these residuals from all spectra in a dataset.

\subsection{Foreground Extinction Correction}

 \begin{figure*}[!t]
 \begin{center}
\includegraphics[width=0.7\paperwidth,clip]{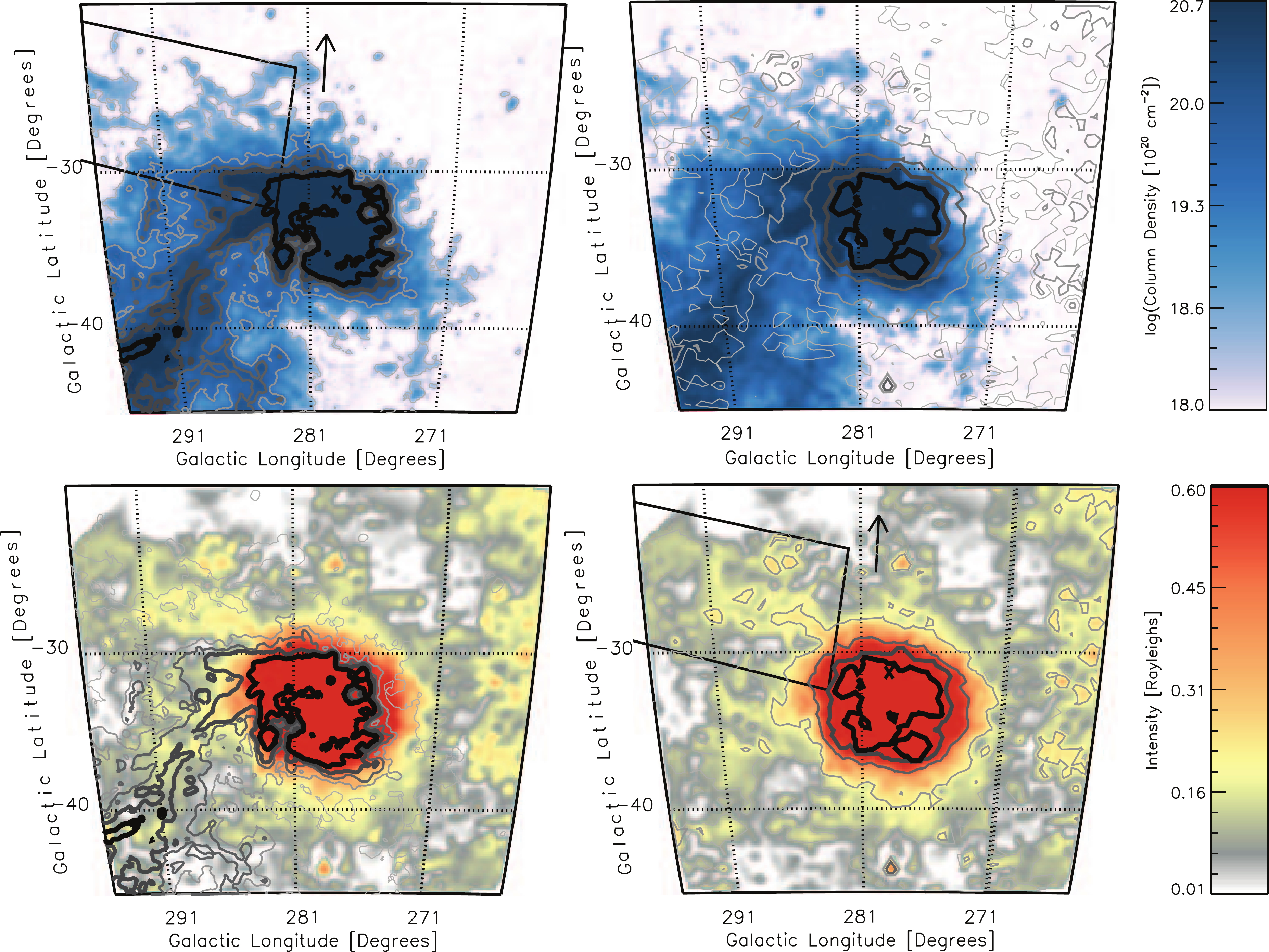}
\caption{Comparison of \hi\ (\emph{top row}) and \ha\ (\emph{bottom row}) emission
maps from the HI4PI survey and the WHAM survey respectively. The left column combines the integrated emission maps with \hi\ contours overlaid, while the right column overlays \ha\ contours. The emission for both maps is integrated over $+150\lesssim\vlsr\lesssim+390~\kms$, which corresponds to $-130\lesssim \vlmcsr \lesssim  +110\,\kms$ at the center of the LMC. The \hi\ scaling is clipped at a column density of $\log{\left(N_{\rm H\textsc{~i}}/\cm^{-2}\right)}=20.7$ to highlight the faint \hi\ emission. The \ha\ emission is clipped at 0.6 R and has not been corrected for MW or internal extinction. The \hi\ contours are at $N_{\rm H\textsc{~i}}=8$, 3.5, 2, 1 and $0.1\times10^{20}$ cm$^{-2}$. The \ha\ contours are at $I_{\rm H\alpha}=10.0$, 2.0, 0.5, 0.2, and 0.1 R. The black box is duplicated from Figure~\ref{fig:Skematic} showing the region identified as the LA. The black X marks the location of 30 Doradus. }
 \label{fig:Dist}
\end{center}
\end{figure*}

Foreground dust from the MW lies between us and the LMC. Therefore, we need to correct the \ha\ intensity for any attenuation associated with the MW. Using the average foreground $R_V$ value of 3.1 as outlined in \citet{Cardelli1989}, we follow the method described in Section 3.4.1 of \citetalias{Barger2013} and apply:

\begin{equation}
I_{H\alpha,corr}=I_{H\alpha,obs}e^{A(H\alpha)/2.5} \label{eq:correction}
\end{equation}
where
\begin{equation}
A(H\alpha)=5.14\times10^{-22} \left\langle \nhi \right\rangle\ \textrm{cm} ^{2}\ \textrm{atoms}^{-1} \ \textrm{mag} \label{eq:exinc}
\end{equation}
and $\left\langle \nhi \right\rangle$ is a measure of the average \hi\ column density over the region of interest.

In this work, the extinction correction is only applied during the mass calculations in Section~\ref{section:Mass}. For foreground MW extinction, we integrate \hi\ spectra from the HI4PI survey \citep{Bekthi2016} over $-100 \leq \vlsr \leq +100$ \kms. We calculated the \textbf{extinction} correction using two methods. The first uses the mean \hi\ for the entire region of interest. The second determines the correction for 0.25\arcdeg\ gridded pixels used in a specific mass calculation method (see Section~\ref{sub:Mass-of-IonizedLMC} for details). In general, we find the range of extinction corrections from these methods increases the observed \iha\ by an average of 10\%--18\% for the main body of the LMC and the LA.

\subsection{LMC Extinction Correction}

\begin{figure*}[!t]
\begin{center}
\includegraphics[trim=0 0 0 0,clip,width=.8\paperwidth]{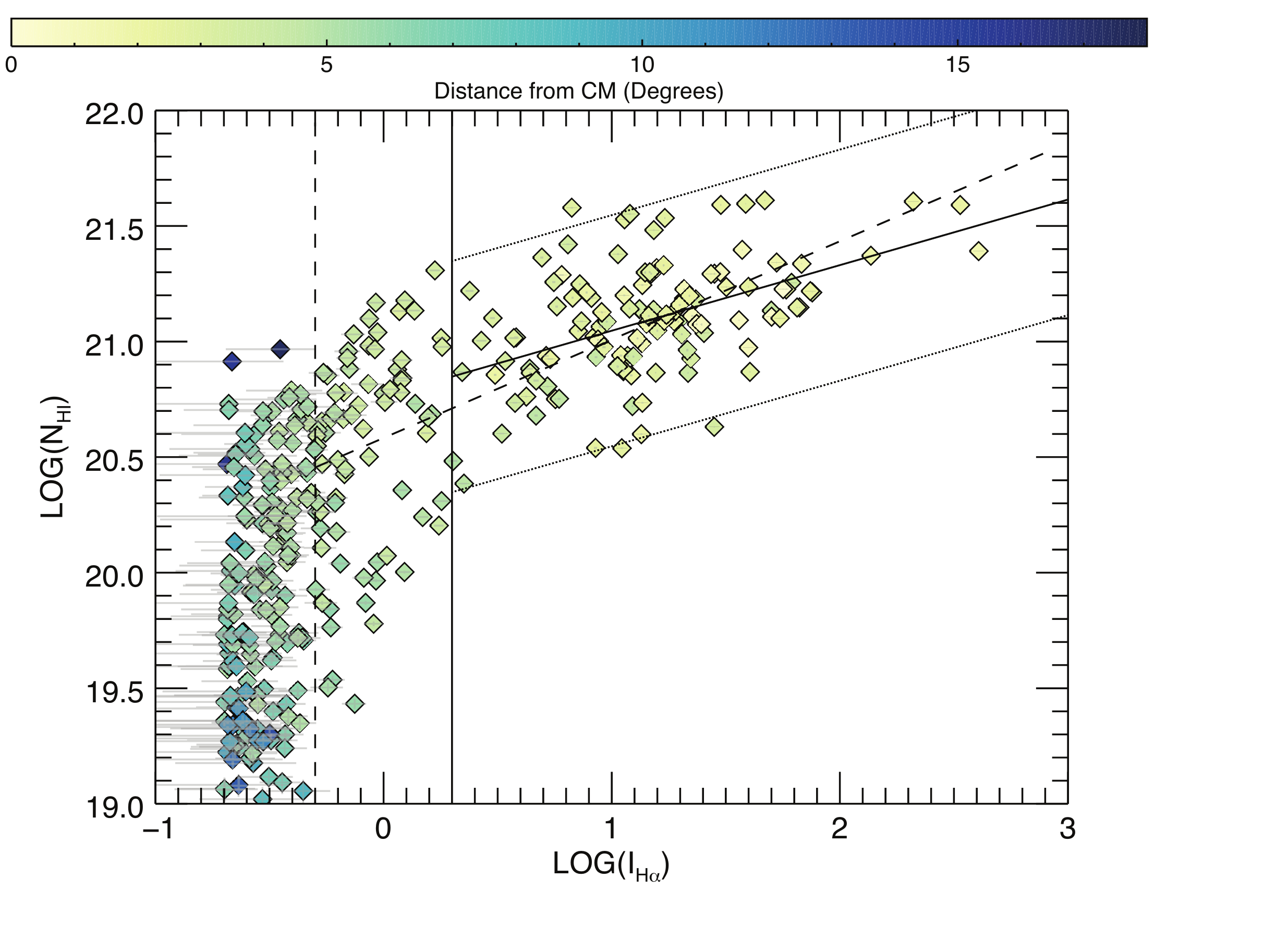}
\par\end{center}
\caption{\ha\ intensity vs. \hi\ column density. The log of the column density of the \hi\ from HI4PI is plotted against the log of the \ha\ intensity in Rayleighs. To match the total area covered by a single WHAM pointing, the \hi\ beams are averaged together to match the WHAM beams. We make a linear fit to all points where log(\iha/\rm R)$\geq -0.3$ and a separate fit to all points where log(\iha/\rm R)$\geq 0.3$. The solid central sloped line fits the points with log(\iha/\rm R)$\geq 0.3$. The lower and upper dotted lines mark the linear fit to the with log(\iha/\rm R) $\pm 0.5$. The dashed vertical line marks the $\log(\iha/\rm R)=-0.3$ cutoff in the sloped dashed line marks the fit. Error bars for both the \nhi\ and \iha\ are in light grey. The colors of the diagram are scaled to the absolute distance from the center of mass of the LMC, $\lb = (279\fdg8,\,-33\fdg5)$ \citep{Kim1998}.\label{fig:HAHIRatio}}
\end{figure*} 

To calculate the internal extinction correction for the LMC, we use values from \citet{Gordon2003} Table 2 and Table 4. Using Table 4, we calculate a $A(\lambda)/A(V)$ for $\lambda=0.656\, \mu$m using their results for wavelengths between $2.198\,\mu$m$\, \ge \lambda \ge 0.440 \,\mu$m. Using a 3rd order polynomial, we have

\begin{equation}
A(\lambda)/A(V)=0.2536\lambda-0.5623\lambda^{2}+0.8468\lambda^{3}-0.1724,
\end{equation}

\noindent where $\lambda$ is in $\mu$m. At \ha\, we then have $ A(\ha) / A(V) = 0.7492 \pm 0.0012 $. We can combine this scaling with $N($\hi$) / A(V)$ from \citet{Gordon2003} Table 2 to estimate $A(\ha)$. This gives us the equation:

\begin{equation}
A\left(H\alpha\right)=2.31\times10^{-22} \left\langle \nhi \right\rangle\ \textrm{cm} ^{2}\ \textrm{atoms}^{-1} \ \textrm{mag}
\end{equation}

We only apply this correction to regions near the center of the LMC that have  integrated $N_{\rm H\textsc{~i}}>1.0\times10^{20}\rm\,cm^{-2}$ over $+150 \le \vlsr \le +390\,\kms$. In these regions, $\iha > 0.5$~R and the average correction is 16\%. Since we do not know the location of the \ha\ emitting gas along the line-of-sight depth with respect to the dust, these extinction-corrected intensities provide an upper limit on \ha\ emission. Due to the low \hi\ column density of the LA, we do not apply a correction for it. Our foreground and internal extinction values are summarized in Table~\ref{tab:Extinction}.

\subsubsection{LMC Velocity Frame}

Due to the large spatial extent of the LMC across the sky, it is useful to adjust velocities to a reference frame with respect to the center of the LMC. In this frame, we can more easily identify extended features that form continuous kinematic structures that are not modified by our local observing geometry. We use the method described in \citetalias{Ciampa2020} to transform the velocity frame of our dataset from LSR to the Large Magellanic Cloud Standard of Rest (LMCSR) given by:

\begin{equation}
\frac{\Delta \rm v_{LMCSR}}{\kms}=262.55 -3.25\left(l-280\arcdeg \right) + 3.66 \left(b-(-33\arcdeg) \right)
\label{eq:VLSREQ}
\end{equation}

\noindent where $l$ and $b$ are Galactic longitude and latitude in degrees. Using Equation \ref{eq:VLSREQ}, $v_{\rm LMCSR}=v_{\rm LSR}-\Delta v_{LMCSR}$. 

\begin{table}[!t]
\begin{center}
\begin{tabular}{ccccc}
\hline 
 & $\rm v_{geo}$  & Wavelength & FWHM & Relative \tabularnewline
Line & [\kms] & [\AA] & [\kms] & Intensity\tabularnewline
\hline 
1 & 182.50 & 6566.80 & 10 & 3.10\tabularnewline
2 & 201.42 & 6567.21 & 10 & 0.90\tabularnewline
3 & 216.90 & 6567.55 & 10 & 1.03\tabularnewline
4 & 238.20 & 6568.02 & 15 & 2.34\tabularnewline
5 & 251.00 & 6568.30 & 10 & 1.84\tabularnewline
6 & 328.00 & 6569.99 & 15 & 1.16\tabularnewline
7 & 337.90 & 6570.21 & 15 & 0.09\tabularnewline
8 & 355.90 & 6570.60 & 15 & 1.30\tabularnewline
9 & 365.70 & 6570.81 & 15 & 1.75 \\ 
\hline
\end{tabular}

\par\end{center}
\caption{Atmospheric Template\label{tab:AtmoLineTable}. This list includes all atmospheric lines observed in the $+150\lesssim\vlsr\lesssim+390~\kms$ window.
The bright OH line at $\rm v_{geo}=+272.44~\kms$ is excluded from our template and fit independently as its strength does not correlate with airmass like these faint lines.}

\end{table}

\section{\ha\  Intensity map\label{sub:IntensityLMC}}
\subsection{Distribution}

In Figure~\ref{fig:Dist}, we present our  non-extinction-corrected total \ha\ intensity and \hi\ column density maps of the LMC. These  maps are integrated over $+150\le \vlsr\le +390\,\kms$. The \ha\ dataset is sensitive to $\iha \backsimeq 10\,{\rm mR}$ for the extended emission structures. The \hi\ dataset was extracted from the HI4PI survey which has a theoretical $5\sigma$ detection limit of $N_{{\rm H}\textsc{~I}}=2.3\times10^{18}\,\text{cm}^{-2}$ \citep{Bekthi2016}.

We compare log(\ha) to log(\nhi) in Figure \ref{fig:HAHIRatio}. We only plot sightlines where the \ha\ emission is above 23\,mR $\left(1\sigma\right)$ and the \hi\ column density is above $10^{19}$~cm$^{-2}$. At $\log(\iha) < -0.3$, there does not appear to be an obvious relationship between the \ha\ intensity and the \hi\ column density. However above that cutoff, there does appears to be a correlation. We tested both ranges of points with the Kendall's Tau correlation test, and found that $\tau =0.52$ for $\iha \ge-0.3\,\rm R$ and $\tau =0.33$ for $\iha \ge0.3\,\rm R$, with the significance level much lower than 0.05. We fit two different slopes to the data using these two data cutoffs:

\noindent $\log{\left(\frac{\iha}{\rm R}\right)} \ge -0.3$:
\begin{equation}
\log{\left(\frac{\nhi}{\rm cm^{-2}}\right)}=20.58 +0.43\times \log{\left(\frac{\iha}{\rm R}\right)} \label{eq:-0.3}
\end{equation}
\noindent $\log{\left(\frac{\iha}{\rm R}\right)} \ge 0.3$: 
\begin{equation}
\log{\left(\frac{\nhi}{\rm cm^{-2}}\right)}=20.76 + 0.28\times \log{\left(\frac{\iha}{\rm R}\right)}
\label{eq:0.3}
\end{equation}

These two fits are shown in Figure \ref{fig:HAHIRatio}. For $\iha\ \ge -0.3\,\rm R$, 11$\%$ of the points fall outside Equation \ref{eq:-0.3} when including an error range of $\pm 0.5$. For $\iha\ \ge 0.3\,\rm R$, only 2$\%$ of the points fall outside Equation \ref{eq:0.3} when including an error range of $\pm 0.5$.  The lines were only fit to the central point of each observation and do not include the errors due to the uneven nature of the error bars in log space. Above $\iha = 0.3\rm\,R$, the errors in both \ha\ and \hi\ are on average 1.3$\%$ of the total \iha\ intensity, and would have had a minor effect on the overall fit.

\subsection{The Leading Arm}

Along with the gas within the LMC, there appears to be an ionized counterpart to the neutral LA (boxed region in Figure~\ref{fig:Skematic}). This emission is more pronounced  when viewing the gas in the \vlmcsr\ frame (Figure~\ref{fig:velcuts}). The LA appears between $-10 \leq \vlmcsr \leq +70$ \kms\ in \hi\ and $-60 \leq \vlmcsr \leq +30$ \kms\ in \ha. The spectra from the regions marked in Figure~\ref{fig:LAMini} are presented in Figure~\ref{fig:LASpectra}. The strongest \ha\ emission lies adjacent to an extended region with $\nhi\gtrsim 0.5\times10^{19}\,\textrm{cm}^{-2}$. Neutral gas is spatially coincident with the ionized hydrogen gas at column densities from $\nhi=\left(0.1-0.5\right)\times10^{19}\,\textrm{cm}^{-2}$ but may not be physically related to each other due to the observed distinct velocities of the two gas components in many locations.

\subsection{\hi\ and \ha\ velocity distribution}
When viewed in the LMCSR frame of reference, the velocity profile of the \hi\ and \ha\ weakly trace each other in the center of the galaxy, around \hi\ contours of $N_{\rm H\textsc{~i}}=10^{20}\ \mathrm{cm}^{-2}$ (Figure~\ref{fig:Vel}). The velocity profile of the \ha\ emission leading to to the LA appears smoothly related to the velocity profiles of the neutral LMC gas. The \ha\ gas does not appear to have the extended high velocity gas that extends out towards the LA. Instead, the lower velocity gas seems dominant in these regions. This difference can be clearly seen in the first moment maps. Unfortunately, the higher velocity is coincident with the OH line at $\rm v_{geo}=+272.44~\kms$ in this region, which prevents us from easily detecting faint, higher velocity \ha\ emission from the LA which coincides with the OH line.

In Figure \ref{fig:VelComp}, we compare the mean velocity of the \ha\ emission to mean($\mathrm{v}_{\mathrm{HI} }$)--mean($\mathrm{v}_{\ha}$) of all points with both an \iha\ $\ge 0.2\,\rm R$ and \nhi $\ge 10^{18}\,\rm cm^{-2}$. We calculate the mean velocity taking the first moment of the spectra between +170 \kms and +390 \kms. We see 28\% of the \ha\ emission has a positive mean velocity, while 72\% has a negative mean velocity. Where the \hi\ has a positive mean velocity, the \ha\ mean velocity has a wider spread. The velocity comparison appears to be weighted towards a more negative mean difference in velocity. This may be biased by the \citepalias{Ciampa2020} observations, which observed the lower velocity LMC winds. Figure \ref{fig:VelCompSpectra} highlights six regions where the absolute difference in velocity between the \ha\ and \hi\ components is greater than 70 \kms. These regions appear on the outer edge of the LMC. The difference in mean velocity may be due to multi-component emission in the region, as seen in Figure \ref{fig:VelCompSpectra} b and d. In other regions, such as Figure \ref{fig:VelCompSpectra} e, the \ha\ and \hi\ components may be related, however the broader nature of the \ha\ emission may shift the mean velocity or the OH line may obscure the true mean velocity of the \ha\ emission.

Since removal of the bright OH line can remove \ha\ emission, we tested how this could bias the \ha\ mean velocity. Adding an extra component to reduced spectra with a velocity width of $15\,\kms$, a center coincident with the OH line, and with intensities of $I_{\rm H\alpha}=0.10$ and 0.23\,R, we found a maximum positive mean velocity shift of 7\,\kms\ for 0.10\,R, and 8\,\kms\ for 0.23\,R. Additionally, even with the shift, the mean \hi\ velocities in the LA are still significantly positive. Any emission underneath the OH line would not shift the \ha\ mean velocity to values similar to  \hi\ mean velocities that are significantly positive.

The separation in velocity space of the ionized and neutral gas may indicate that, despite the spatial coincidence from our perspective, the gas components are physically separated. This could indicate separate origins for the ionized and neutral gas or additional mechanisms affecting different regions of gas. A multi-component analysis in a future work may reveal a more accurate relationship between the neutral and ionized gas components.
 
 \begin{figure*}[!htbp]
   \begin{centering}
       \includegraphics[width=0.95\textwidth]{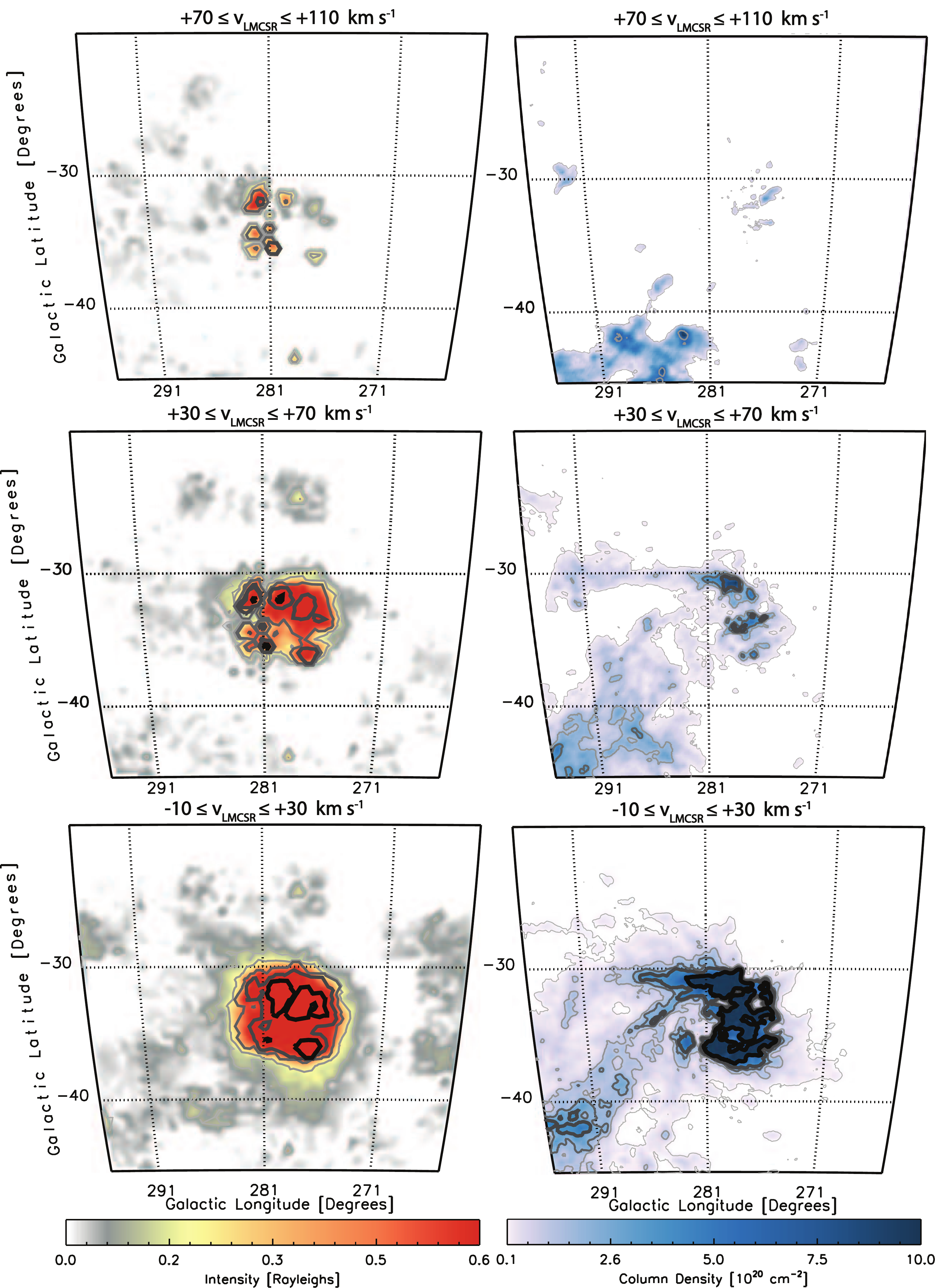}
     \end{centering}
     \caption{Velocity Slice Maps. Comparison of 40-\kms-wide velocity windows covering $-130\leq\vlmcsr\leq+110\,\kms$. Velocity ranges are noted at the top of each figure. The left column shows the WHAM \ha\ emission and the right column shows the \hi\ emission from HI4PI.}
     \label{fig:velcuts}
\end{figure*}

     \pagebreak
    
\begin{figure*}[!htbp]
    \figurenum{\ref{fig:velcuts}}
     \begin{centering}
       \includegraphics[width=0.95\textwidth]{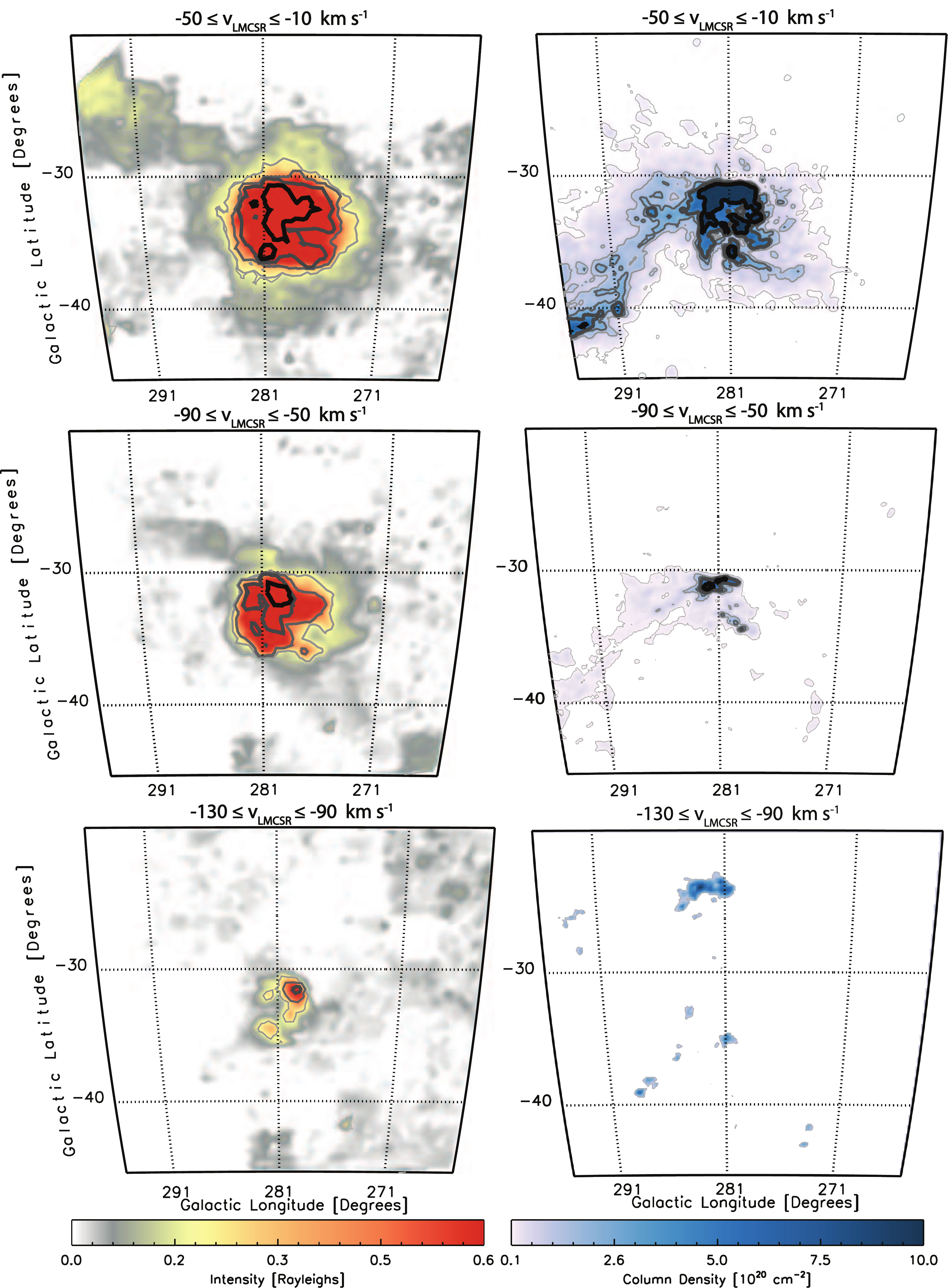}
    \end{centering}
    \caption{\emph{(continued)}}
\end{figure*}

\section{$H^0$ and $H^+$  Mass \label{section:Mass}}

We calculate the mass of ionizing gas following the methods presented in \citet{Hill2009}, \citetalias{Barger2013} and \citetalias{Smart2019}. Emission measure is related to the integral of the path length times the square of the electron density, $\mathrm{EM}=\int n_{e}^{2}dl$. As the \ha\ emission is proportional to the recombination rate, $\iha=(4\pi)^{-1}\int \alpha_{B}(T)\epsilon_{\ha}(T)n_{e}n_{p}dl_{H^{+}}$, we can combine the two to rewrite the emission measure as follows:

\begin{equation}
EM=2.75\left(\frac{T}{10^{4}\,\textrm{K}}\right)^{0.924}
\left(\frac{\iha}{\textrm{R}}\right)
\textrm{cm} ^{-6}\,\textrm{pc}  \label{eq:EMEQ}
\end{equation}

where $n_{p} \approx n_{e}$ and the probability that the recombination will produce \ha\ emission is $\epsilon_{\ha}\left( T \right) \approx 0.46 \left(T/10^{4} \mathrm{K} \right)^{-0.118}$. The recombination rate assumes the gas is optically thick to ionizing photons, $\alpha_B = 2.584 \times 10^{-13}\left(T/10^{4} \mathrm{K} \right)^{-0.806} \mathrm{cm}^{3}\,\mathrm{s}^{-1}$. The integrated line-of-sight $dl$ is assumed to be the $H_{+}$ line-of-sight depth, and T is kept constant at $10^{4}$ K, the characteristic temperature of the WIM. All \iha measurements are from the WHAM dataset.

Since \ha\ traces $n_e^2$ and not $n_e$ directly, we must make assumptions for either the line-of-sight depths or the electron density to estimate an ionized mass. The line-of-sight depth and the electron density are related by the previous equation $EM=\int n_{e}^2dl$, thus we can use this relation and Equation \ref{eq:EMEQ} to calculate whichever value is not assumed:

\begin{equation}
\L_{H+} =2.75\left(\frac{T}{10^{4}\,\textrm{K}}\right)^{0.942}
\left(\frac{\iha}{\textrm{R}}\right)
\left(\frac{n_{e}}{\textrm{cm} ^{-3}}\right)^{-2}\,\textrm{pc}  \label{eq:LineofSight}
\end{equation}

Following \citet{Hill2009}, we assume that the mass of the region can be calculated using $M_{H_{+}}=1.4m_{H}n_{e}D^{2}\Omega L_{H^{+}}$. Combining Equation \ref{eq:EMEQ} with the equation for mass, the resulting mass in each beam is then: 
\begin{equation}
\left(\frac{M_{\textrm{H}^{+}}}{\msun}\right) = 8.65\times10^{7}\,\Omega 
\left(\frac{D}{50 \text{kpc}}\right)^{2}\left(\frac{EM}{\textrm{cm}^{-6} \textrm{pc}}\right)\left(\frac{n_{e}}{\text{cm}^{-3}}\right)^{-1}\label{eq:Mass}
\end{equation}

where $\Omega$ is the solid angle over which the mass is calculated. In this paper, $\Omega$ is either a $0\fdg25$ pixel in a re-sampled image when calculating individual sight-lines or the total solid angle observed when averaging the entire region. The factor of 1.4 in the mass calculation accounts for helium \citep{Hill2009,Barger2013}. D is the distance to the LMC, assumed to be 50 kpc unless specified otherwise. 

\subsection{Mass of Ionized Gas\label{sub:Mass-of-IonizedLMC}}

 \begin{figure*}[!t]
\begin{centering}
\includegraphics[clip,width=0.7\paperwidth]{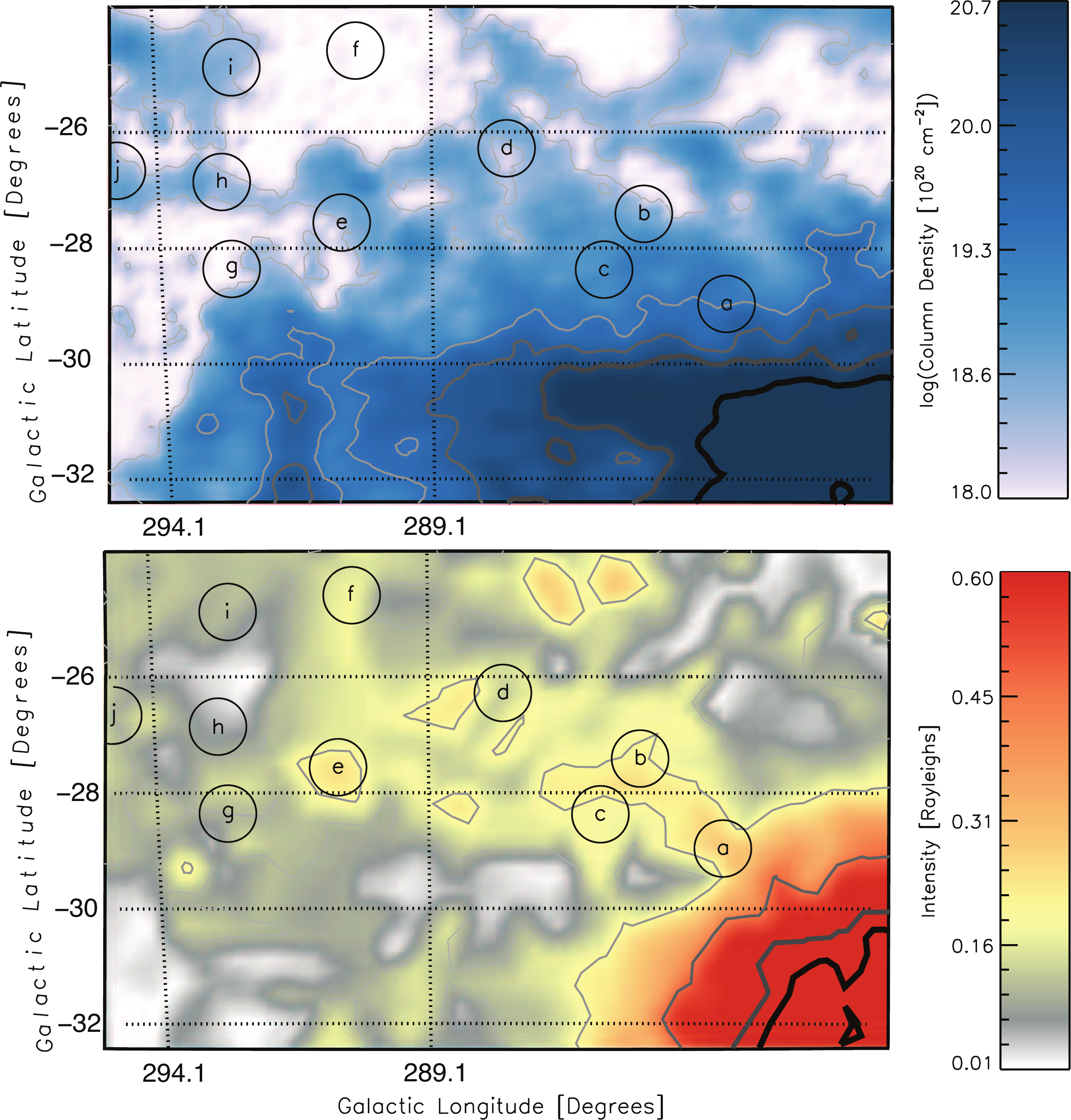}\tabularnewline
\par\end{centering}
\caption{Comparison of \hi\ and \ha\ in the LA1.1 region integrated over $+150 \le \vlsr \le +390\,\kms$. Black circles mark the location of spectra plotted in Figure~\ref{fig:LASpectra} and the size of the WHAM beam. \label{fig:LAMini}}
\end{figure*}

\begin{figure*}[!htbp]
\begin{centering}
\includegraphics[clip,width=0.7\paperwidth]{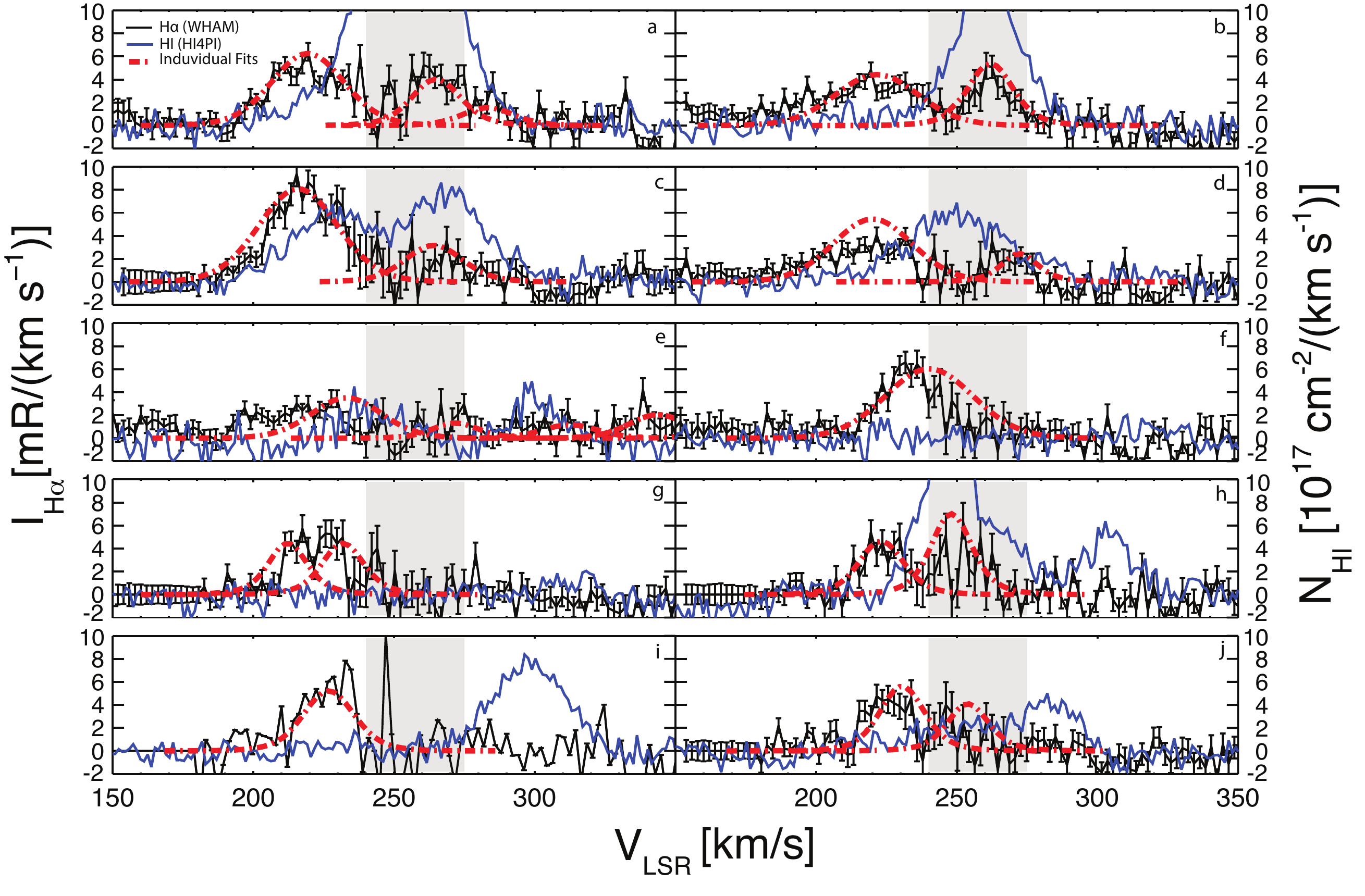}\tabularnewline
\par\end{centering}
\caption{Kinematics of six selected locations within the Leading Arm. Black lines show the \ha\ emission observed by WHAM, while blue lines trace \hi\ emission from the HI4PI survey. The red dashed lines show individual component fits to the \ha\ spectra. The location of each sightline is marked in Figure~\ref{fig:LAMini}. The grey region marks the area where the OH line is present. \label{fig:LASpectra}}
\end{figure*}

\begin{figure}[!tbp]
\begin{centering}
\includegraphics[width=0.9\columnwidth,clip]{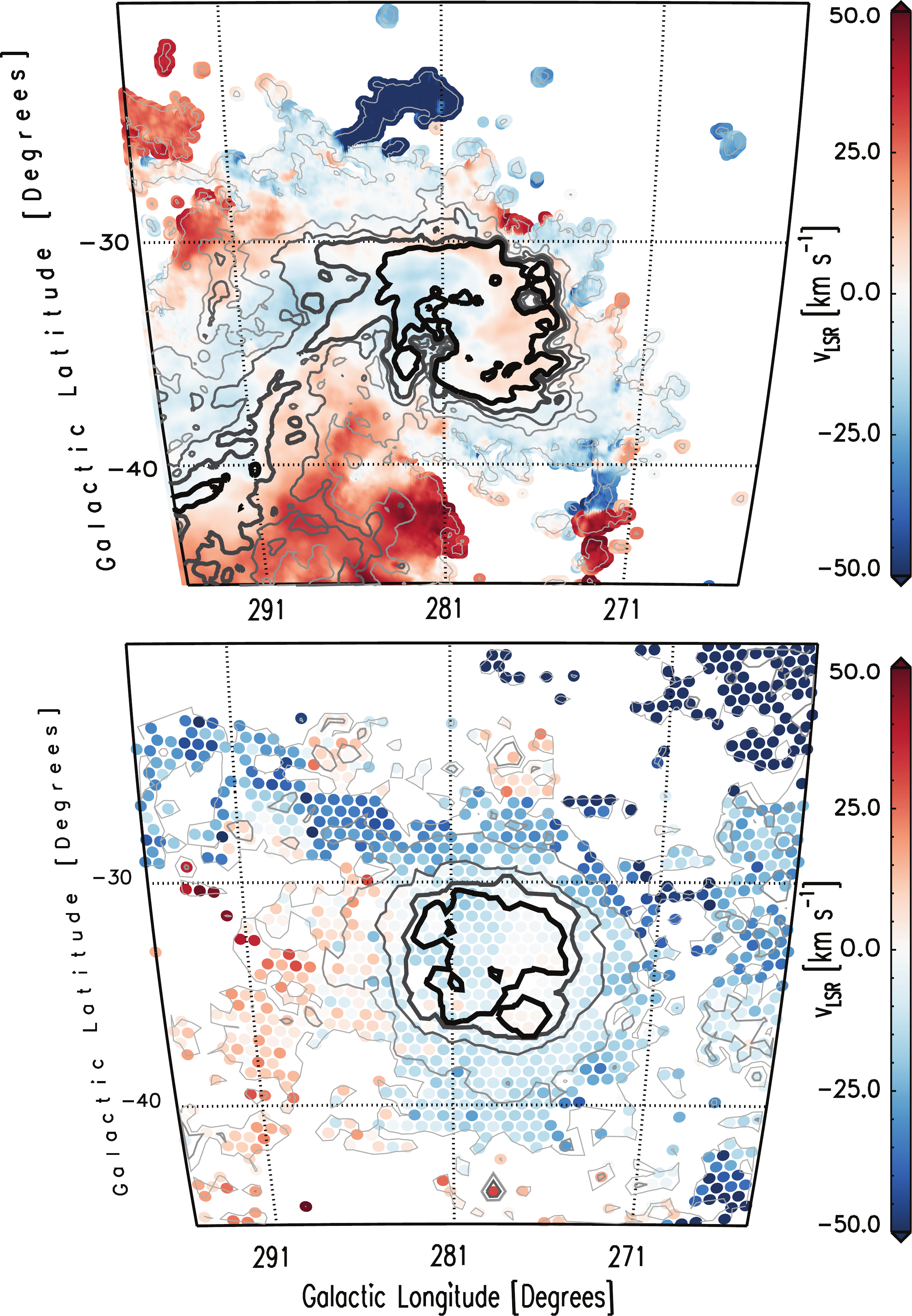}
\par\end{centering}
\centering{}
\caption{First Moment Map. The mean LMCSR velocities of the gas traced by the \hi\ (\emph{top}) and the \ha\ (\emph{bottom}). The rotation of the ionized gas in the center of the galaxy weakly mirrors the motion of the neutral gas. Only pointings with $\iha = 10\,\rm mR$ or $N_{\rm H\textsc{~i}}\ge 10^{18}\rm cm^{-2}$ are plotted.   \label{fig:Vel}}
\end{figure} 

\begin{figure*}[!t]
\begin{center}
\includegraphics[trim=0cm 0 0 0,clip,width=.7\paperwidth]{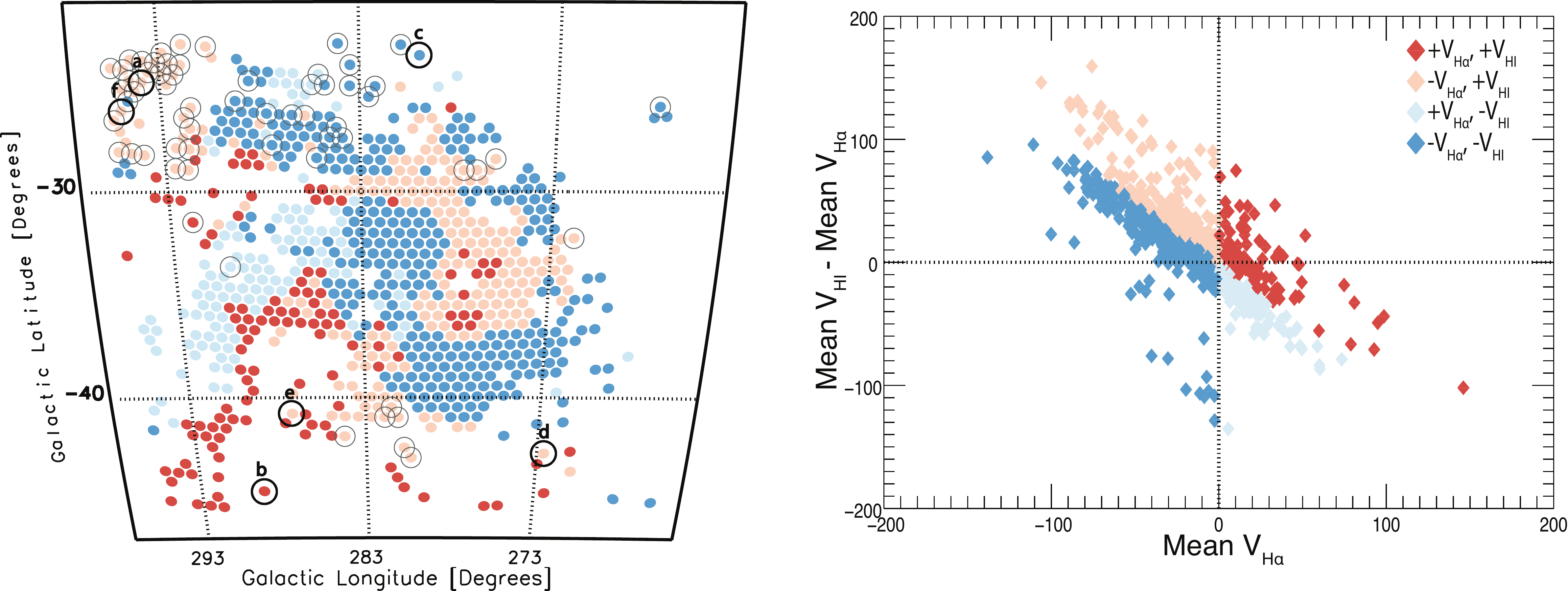}
\par\end{center}
\centering{}
\caption{Mean velocity comparison. On the right we plot the mean \ha\ \vlmcsr\ of all points with both \ha\ and \hi\ emission above $I_{\rm H\alpha}>0.1\,R$ and $N_{\rm H\textsc{~i}}10^{19}\rm\,cm^{-2}$ vs. the difference between the difference between the mean \hi\ velocity and the mean \ha\ velocity. On the left we have a map of the points with the same color coding as the right plot. Light grey circles mark all sightlines where the absolute difference in mean \vlsr\ is greater than 70\,\kms. Of these sightlines, we selected six representative locations---marked with black circles and labeled a--f---and plotted their corresponding \hi\ and \ha\ spectra in Figure \ref{fig:VelCompSpectra}. \label{fig:VelComp}}
\end{figure*} 

\begin{figure*}[!t]
\begin{center}
\includegraphics[trim=0cm 0 0 0,clip,width=.8\paperwidth]{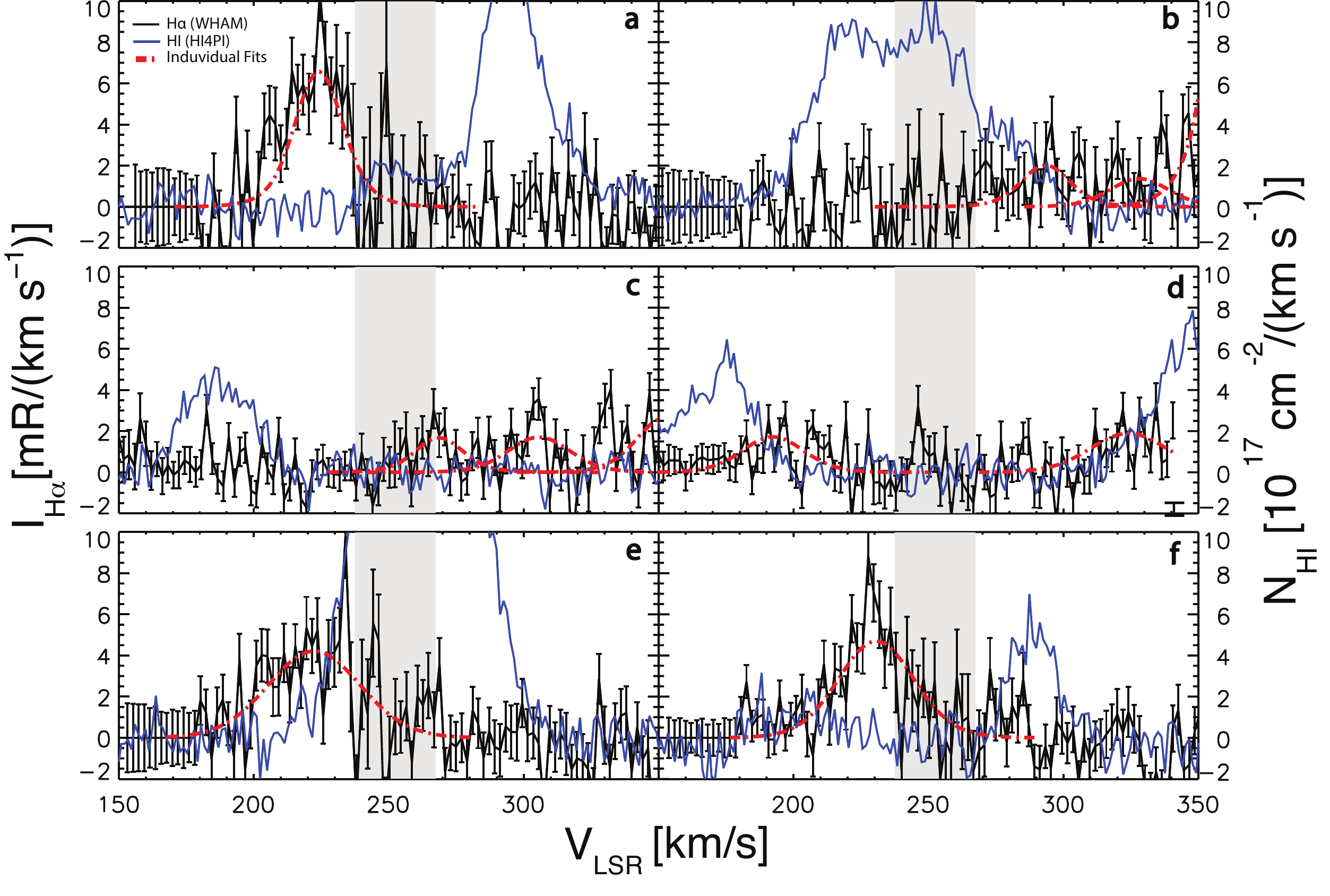}
\par\end{center}
\centering{}
\caption{Mean velocity spectra comparison. We selected the \hi\ and \ha\ spectra from sightlines that have an absolute mean velocities differences greater than 70 \kms. Black lines show the \ha\ emission observed by WHAM, while blue lines trace \hi\ emission from the HI4PI survey. The red dashed lines show individual component fits to the \ha\ spectra. The location of each sightline is marked in Figure~\ref{fig:VelComp}. The grey region marks the area where the OH line is present.\label{fig:VelCompSpectra}}
\end{figure*} 

In this work, we estimate the mass of the LMC three different ways. As the electron density and the line-of-sight depths of the gas is unknown, we must make generalized assumptions about the distribution of the ionized gas in relation to the neutral components.

We first describe a mass estimate where \hi\ measurements from HI4PI are used to constrain the electron density (\S\ref{sub:neno}). This assumes the ionized gas exists in a skin around the neutral gas. 

We then describe two different mass estimates where we assume the neutral and ionized gas are well mixed and share the same the line-of-sight depth. These two scenarios make different assumptions about the geometry of the galaxy. Our second method assumes a cylindrical geometry, which treats the gas as a uniform, cylindrical slab which we view from the top down. 

The third method assumes the gas within the LMC exists in an ellipsoid centered at the LMC's kinematic center. 

In each of these methods, the line-of-sight depth of the LMC's gas components is not known. We use the maximum line-of-sight depth of the stellar component measured in  SMASH survey (7kpc \citep{Choi2018}) as a proxy (\S\S\ \ref{sub:neno}, \ref{sub:cylinder} \& \ref{sub:ellipse}) to estimate the upper limit of the assumed line-of-sight depths for either the neutral or the hydrogen gas. We also use a 3 kpc depth, which assumes the gas component has a smaller line-of-sight depth than the stellar component. For the LA, we constrain the line-of-sight depth to 1.5 kpc. The assumption of 1.5 kpc comes from the observed spatial extent of the HI features in the LA area on the plane of the sky. It is reasonable to expect that the features have comparable extents along the line-of-sight as they do in the plane of the sky.

For each method described below, we focus on two regions, the LMC and the LA (Figure~\ref{fig:Skematic}). The LA region is defined as LA1.1 in \citet{Venzmer2012}. We use a modified version of their bounds, \lb $=(295,-23.83$), ($281.78,-25.51$),($283.29,-32.41$), ($295,-29.51$), to accommodate the boundaries of our observations. The area covered by the LMC and the LA regions is 171.3 and 78.5 degrees$^{2}$ respectively. All mass calculations were determined by integrating emission over $+150\leq \vlsr \leq+390\,\kms$. Any directions with \ha\ intensities below 25\,mR are excluded when calculating the mass using individual sightlines. To match the HI4PI observations to the WHAM beams, we take the HI4PI column density observations contained in each beam and average them together. We adopt a distance to the LMC of $D=50\,{\rm kpc}$ \citep{Walker2012,Pietrzyski2013}, although the ellipsoid scenario in Section~\ref{sub:ellipse}  uses the inclination, $i = 25\fdg86$ \citep{Choi2018}, to vary D. Values for each region can be found in Table \ref{tab:ModelTable}. The resulting atomic gas ratios, \added{$M_\mathrm{ionized}/M_\mathrm{ionized+neutral}$, are listed in Table~\ref{tab:IonizationTable}.}

\subsubsection{$n_e = n_0$\label{sub:neno}}
The first mass calculation uses the averaged \hi\ column density to estimate the electron density of the region. We make the assumption that the ionized gas lies in a skin around the neutral gas. This assumption results in two possible relations between the ionized and neutral gas.  If the neutral and ionized components are separated, but are in pressure equilibrium, then the electron density of an
ionized skin would equal half the neutral hydrogen density, $n_e = n_0/2$ 
\citep{Hill2009}. However, if they are not in pressure equilibrium they would equal $n_e = n_0$. For the LMC, the recombination time ($\mathbf{\sim}$1 Myr) is much shorter than the sound crossing time (a few
hundred Myr). Thus we only use $n_e = n_0$, assuming they are not in pressure equilibrium.

Using $n_e = n_0$ , we average the HI4PI column densities together for each regions marked in Figure \ref{fig:Skematic} and and assume a line-of-sight depth of 3 kpc for the neutral gas. We use the resulting average and assumed depth to calculate a number density for the neutral gas, and use this as an estimate for the electron density. With the average electron density we can calculate the assumed line-of-sight depth using Equation \ref{eq:LineofSight}. We then use Equation \ref{eq:Mass} to calculate the total ionized mass based on the averaged electron density. In this method, $\Omega$ is assumed to be the defined angular area of the region. Using the maximum stellar depth produces an unrealistic upper mass estimate for the ionized gas, and is excluded.  For the LMC, we find $M_{\rm ionized}\approx1.6\times10^{9}\,\msun$  with an ionized line-of-sight depth of 5.1 kpc. For the LA we find a mass of $M_{\rm ionized}\approx54\times10^{6}\,\msun$.

\subsubsection{Cylindrical Geometry\label{sub:cylinder}}
The second scenario treats the gas in the galaxy as a cylindrical slab along our line-of-sight. We assume $L_{\mathrm{H}^+} = L_{\mathrm{H}^0}$, with the neutral and ionized components occupying the same volume along our line-of-sight. The intensity is then averaged over the region and treated as if it is in a single, flat cylinder viewed from above with the same line-of-sight everywhere(3 kpc or 7 kpc) and uses the total angular area of the defined region as $\Omega$. We then use Equation \ref{eq:LineofSight} to find the electron density. The two different line-of-sight depths result in a mass range of $M_{\rm ionized}\approx\left(1.2-1.8\right)\times10^{9}\,\msun$ respectively. In the LA region we find a mass of $M_{\rm ionized}\approx62\times10^{6}\,\msun$.

\subsubsection{Ellipsoidal Geometry\label{sub:ellipse}}
The ellipsoid scenario is similar to our cylindrical scenario, however instead of averaging the region we apply the mass calculation and the extinction correction to individual 0\fdg25 pixels, allowing a direct comparison between \hi\ and \ha\ emission in each location. In this scenario, we model the shape of the LMC as a simple ellipsoid, which is then used to estimate the line-of-sight depth of the gas. We centered our ellipsoid around the \hi\ kinematic center at $\lb = (279\fdg8,\,-33\fdg5)$ from \citet{Kim1998}. The projected ellipse can be seen in Figure~\ref{fig:Skematic}, with a semi-major axis of 7 kpc and a semi-minor axis of 6 kpc, and a position angle $\theta = 149.23$ degrees \citep{Choi2018}.  We then assume that the maximum line-of-sight depth of the ellipse is similar to the stellar line-of-sight depths or are half the maximum depth (7 kpc or 3 kpc). We then use vary the line-of-sight depth as we move away from the center.

In both cases, if we simply use the equation for an ellipsoid the outer regions of the LMC would fall to sub kpc line-of-sight depths. We chose to constrain the minimum depth for the model to be no less than 1.5 kpc. In addition to the varying line-of-sight we vary the distance, D  from equation \ref{eq:Mass}. The distance to the center of the LMC is assumed to be $\textrm{D} = 50$ at the center of the galaxy. In this scenario we allow the inclination of the galaxy to change D according to the following equation:

\begin{equation}
D = 50\,\textrm{kpc} + d \times tan \left(\theta\right)
\end{equation}

where $\theta = 25\deg$ and d is the positive or negative distance of the pixel in kpc from the center of the LMC along the semi-major axis, with the positive direction towards the NE side of the galaxy and the negative towards the SW. The inclusion of inclination in the model had a $\sim1\%$ effect on the mass, due to the inclination having the largest effect on the regions furthest from the center of the galaxy where there is less gas and the symmetry of the inclination minimizing its overall effect. 

Extinction corrections are estimated for each pixel in a 0\fdg25 grid. We also limit our mass calculations to sightlines with \ha\ emission above $I_{\rm H\alpha}>25\,\rm mR$, except in the case of the LA where we also include observations down to 10 mR to account for the significant amount of continuous faint emission in the region. Using this model, we find ionized hydrogen that is traced by the \ha\ emission in the LMC to be $M_{\rm ionized}\approx6\times10^{8}\,\msun$ using a depth of 3~kpc and $M_{\rm ionized} \approx10\times10^{8}\,\msun$ using a depth of 7~kpc.

In the LA, the depth is constrained to a flat slab with a dept of 1.5\,kpc and the distance is kept at 50\,kpc similar to the cylindrical calculations. However, we use the individual sight-line calculations instead of the averaged intensity and the same intensity cut-offs we use in the ellipsoidal scenario. The resulting ionized gas mass for the LA is $\left(12-46\right)\times10^{6}\,\msun$ for 25\,mR and 10\,mR  intensity cutoffs.

Each of these scenarios treats the gas as if it is homogeneously distributed. They ignore the possibility of a clumpy WIM or a diffuse halo with a thick, dense disk of ionized gas. This likely leads to an overestimate of the total ionized gas mass. Thus, our estimates should be treated as upper limits to the total gas mass and ionization fraction. Future work will investigate how varying the gas distribution effects the total mass calculated.

\begin{table*}[!t]
\caption{Neutral and Ionized Properties\label{tab:ModelTable} }
\begin{center}
\scriptsize
\begin{tabular}{ccccccccccc}
\hline 
\hline 
{Region} & & \multicolumn{6}{c}{Neutral Properties} &  &\tabularnewline
& \cline{2-7} 
 & & $\Omega$& $ \log \left\langle \nhi \right\rangle\tablenotemark{a} $ & $M_{\mathrm{H}^{0}}$\tablenotemark{b} & $M_{\mathrm{H}^{0}}$\tablenotemark{c} & $L_{\mathrm{H}^{0}}$ & $\log \left\langle n_0 \right\rangle $ & \tabularnewline
 & & $\deg^{2}$ &[cm$^{-2}]$ & [10$^{6}$ \msun] & [10$^{6}$ \msun] &[kpc] & [cm$^{-3}$] &\tabularnewline
\hline 
LMC & & 171.3 -- 170.0 &20.7 & 850 & 760 & 3.0--7.0\tablenotemark{d} &-1.6 -- -1.2 & \tabularnewline
Leading Arm 1.1 & & 78.5 --79.8 & 20.0 & 71 & 68& 1.5 & -1.6 &  \tabularnewline
\hline 
\hline 
{Region} & &  \multicolumn{6}{c}{Ionized Properties}  \tabularnewline
& &   \multicolumn{2}{c}{Ionized Skin} & & &  \multicolumn{2}{c}{Cylinder} & & \multicolumn{2}{c}{Ellipsoid}   \tabularnewline
\cline{3-5} \cline{6-7} \cline{7-11} 
 & $\log\left\langle EM\right\rangle $ & $\log \left\langle n_{e}\right \rangle $\tablenotemark{e} & $L_{\mathrm{H}^{+}}$ & $M_{\mathrm{H}^{+}}$ & $\log \left\langle n_{e} \right \rangle $ &   $L_{\mathrm{H}^{+}}$ & $M_{\mathrm{H}^{+}}$ & $\log \left\langle n_{e} \right \rangle\tablenotemark{d}$ & $L_{\mathrm{H}^{+}}$ & $M_{\mathrm{H}^{+}}$  \tabularnewline
 & [cm$^{-6}$ pc] & [$ \mathrm{cm}^{-3}$] &[kpc] & [$10^{6}$ \msun] & [$\mathrm{cm}^{-3}$] &[kpc] & [$10^{6}$ \msun] & [$\mathrm{cm}^{-3}$] &[kpc] & [$10^{6}$ \msun] \tabularnewline
\hline 
LMC & 1.4--1.4 & -1.6 & 5.1 & 1600 & -1.1 -- -1.2 & 3.0 -- 7.0 & 1200--1800 & -0.9 -- -1.1 &3.0 -- 7.0\tablenotemark{f} & 600--1000 \tabularnewline
Leading Arm 1.1\tablenotemark{g} & -0.2-- -0.3 & -1.6 &0.8 & 54  &  -1.7 & 1.5 & 62 & -1.7 & 1.5 & 12--46 \tabularnewline
\hline 
\end{tabular}
\end{center}

\tablenotetext{a}{Average H\textsc{~i} column density for the region calculated from HI4PI.}

\tablenotetext{b}{Mass range calculated using the ionized skin (\S\ref{sub:neno}) and cylinder (\S\ref{sub:cylinder}) scenarios.}

\tablenotetext{c}{Mass range calculated using the ellipsoid scenario (\S\ref{sub:ellipse}) only.}

\tablenotetext{d}{$n_{e}$ is calculated over the respective line-of-sight for each scenario.}

\tablenotetext{e}{ Only uses an assumed line-of-sight for  $\log \left\langle n_0 \right\rangle $ of 3 kpc for the LMC. } 

\tablenotetext{f}{Maximum line-of-sight depth at the center of the LMC for the ellipsoid scenario. 3\,kpc and 7\,kpc maximum depths are from \citet{Choi2018}. Minimum depth used is 1.5\,kpc.}

\tablenotetext{g}{Mass range calculated with an emission cut-off of $I_{\rm H\alpha}=25\,\rm mR$ and 10\,mR.}

\end{table*}

\begin{table*}[!t]
\caption{Ionization Mass Fraction: $M_\mathrm{ionized}/M_\mathrm{ionized+neutral}$\label{tab:IonizationTable} }
\begin{center}
\begin{tabular}{cccc}
\hline 
\hline 
 & Ionized Skin ($n_e = n_0$) & Cylinder& Ellipsoid \tabularnewline
\hline 
 LMC & $65\%$ & 58--68\%& $46$--$58\%$ \tabularnewline
  LA & 43\% & 47\% & $15$--$40\%$ \tabularnewline
\hline 
\end{tabular}
\end{center}
\end{table*}

\section{Discussion\label{sub:DiscussionLMC}}
 
\subsection{Leading Arm}

LA1.1 ( see Figure~\ref{fig:Skematic}) is the closest region of a larger section of the LA system, LA1 \citep{Nidever2008,Venzmer2012}. Each sightline marked in Figure~\ref{fig:LAMini} and Figure~\ref{fig:LASpectra} contains \hi\ and \ha\ emission spanning similar velocities, including several regions where both the \hi\ and \ha\ appear to have multiple components. The first moment map in \ha\ in this region appears to be weighted towards lower velocities compared to the \hi\ gas in the region (Figure~\ref{fig:Vel} and Figure \ref{fig:VelComp}). However, the higher velocities can overlap with the bright OH sky line where faint \ha\ emission is difficult to trace. As a result, it is difficult to determine if the \ha\ emission is genuinely multi-component or a single, broad feature.
 
 The COS/UVES survey contains one absorption target present in the LA1.1 region. PKS0637 lies within the region defined as LA1.1 by \citep{Venzmer2012}. Their modeling finds an ionization fraction of 59\% along the line-of-sight coincident with $\log{\left(\nhi/\cm^{-2}\right)}=19.29$. Although our determination of the ionization fraction is less constrained for the LA (16--56\%), it is consistent with their estimate. 
Along $b=-30\arcdeg$ \hi\ column density falls of quickly from \nhi\ $ \ge 8\times10^{20}\rm\,cm^{-2}$ down to \nhi\ $ \le 10^{20}\rm\,cm^{-2}$. In contrast, \ha\ emission appears to connect  continuously from the LMC to the LA, both spatially and in velocity.

In \citet{Nidever2008}, the radial velocities of the LA complexes were measured and found to be similar to those of the LMC. They initially identified gaps between the separate components of the LA1, however deeper observations by the \hi\ Parkes All-Sky Survey (HIPASS) show the region to be continuous in neutral gas. In the \ha\ emission, we have identified what appears to be a continuous filament of ionized gas that extends out towards the more distant \hi\ components of LA1. The filament of gas lies at latitudes $b> -29\arcdeg$ where $\log{\left(\nhi/\rm cm^{-2}\right)} < 20$ (Figure~\ref{fig:LAMini}).

Contrary to \citet{Nidever2008}, a study by \citet{Putman1998} argues for the SMC as an origin to LA1. This claim is based on an \hi\ feature originating from the SMC and smoothly extending to the base of LA1.1 in velocity space (see their Figures 1 and 2). In addition, \citet{Staveley-Smith2003} suggest that the gas from the LMC is merely leaking into the LA and not a dominant contributor.

While our results for the kinematics and spatial extent of the ionized gas in this region match gas velocities from the LMC, the spatial extent of our observations is limited. However, in our observed region, the velocity and spatial extent of the gas appear to connect smoothly with the LMC, suggesting the LMC is the origin of the ionized component of LA1.1  Additionally, some sightlines (Figure \ref{fig:Vel}) with velocities that match the Bridge region extend above the LMC. This may indicate some material within the region is associated with the SMC or Bridge. Further observations of the full extent of the LA to find \ha\ emission present in the gaps in \hi\ column density will test the hypothesis that LA1 is a continuous structure that originated from LMC.

 \subsection{Magellanic System}
 
 The results of this LMC study extend  our WHAM survey of the Magellanic System. \citetalias{Smart2019} find a gas ionization fraction of  42\%--47\% associated with the central region of the SMC and an \textbf{ionized gas} mass range of $M_{\rm ionized}\approx\left(6-7.5\right)\times10^{8}\,\msun$ \textbf{ and a neutral gas mass range of }$M_{\rm ionized}\approx\left(8.4-8.7\right)\times10^{8}$. Our results listed in Table \ref{tab:IonizationTable} show a significantly higher ionization fraction for the LMC using the ionized skin and cylinder method. Using the ellipsoidal method, we find a similar ionization fraction found for the SMC when assuming a maximum thickness of 3\,kpc.
 
 Like the SMC, the LMC ionized gas follows neutral gas rotation within the center of the galaxy where $\log{\left(\nhi/\rm cm^{-2}\right)} > 20.7$, though it appears to be a weak correlation. Similar to the SMC, there appear to be regions of gas which follow the velocity trends seen in the center of the galaxy and extend out into the halo of the galaxy. The LMC has the LA feature, and the SMC has an ionized filament that extends out of the galaxy which smoothly connects to the central rotating ionized gas in velocity space \citepalias{Smart2019}. 
 
Similar to the two galaxies, the Bridge also appears significantly ionized. The ionized gas mass for the Bridge ranges from $M_{\rm ionized}=\left(0.7-1.7\right)\times10^{8}\,\msun$ with a neutral gas mass of $M_{\rm neutral}=3.3\times10^{8}\,\msun$ \citepalias{Barger2013}. Like the ionized gas in the SMC and LMC, the \ha\ emission associated with the Bridge extends beyond the boundaries of the neutral gas.
 
 Combined, these three studies have discovered a total of $M_{\rm ionized}=\left(1.3-2.6\right)\times 10^{9}\,\msun$ in diffuse ionized gas associated with the Magellanic System. Compared to the total neutral gas mass of $M_{\rm neutral}=\left(1.9-2.1\right)\times10^{9}\,\msun$ from these studies, the gas in these galaxies and their extended environment is significantly ionized. \citet{Hopkins2012} modeled star-forming galaxies of varying mass ranges and investigated the warm $\left(2000 \mathrm{K}  < T < 4 \times 10^{5} \mathrm{K} \right)$  gas to cold ($T < 2000\,\rm K$) gas ratios of the star-forming disks. Their models found a warm-to-cold gas ratio of 0.5-0.7 for MW and SMC-like galaxies. Another study by \citet{Dobbs2018} investigated the ISM in star-forming spiral galaxies. They split their gas fractions into cold gas $\left(\mathrm{T} < 150\, \mathrm{K}\right)$, intermediate gas $\left(150 < \mathrm{T}< 5000\, \mathrm{K}\right)$, and warm gas $\left(\mathrm{T} > 5000\,\mathrm{K}\right)$ and find the total fraction of gas in each state is roughly equal. While neither of these studies is directly analogous to our observations, we find comparable warm-to-cold gas ratios.

\subsection{The Magellanic Corona}
 
 Many models of the MSys fall short of replicating both the MS and LA in extent and morphology, thus investigations into the extended material are important for informing future models. A recent paper by \citet{Lucchini2020} investigates the impact of adding an envelope of warm-hot coronal gas surrounding the two galaxies, called the Magellanic Corona. The Magellanic Corona is defined as a halo gas at a transition temperature of $\sim10^{5}$ K surrounding the LMC with a mass of $~3\times10^{9}\,\msun$ extending over the virial radius, 100 kpc. 

The warm ionized gas studied here is at $10^{4}$~K \citep{Hoyle&Ellis1963,Haffner2003} and is not directly comparable to the warm-hot corona defined in the \citet{Lucchini2020} models. At $10^{4}$ K, the relation between the measured \ha\ intensity and the emission measure scales according to Equation \ref{eq:EMEQ}.  For temperatures above $2.6\times10^{4}$ K, the \ha\ intensity is highly sensitive to temperature. From \citet{Draine2011},

\begin{equation}
\begin{split}
EM=2.77\,\textrm{cm} ^{-6}\,\textrm{pc}
\left(\frac{\iha}{R}\right)
\left(\epsilon_{\textrm{b}}\right)^{-1}\\
\times T_{4}^{\left(0.942+0.031 \,\textrm{ln}T_{4}\right)}
\end{split}
\end{equation}  

\noindent where $T_{4}=T_{e}/10^{4}$ K and $\epsilon_{b}$ is the beam dilution factor. Assuming $\epsilon_{b}=1$ and $T_{e}=10^{5}$ K, 25 mR would result in $EM=0.71 \,\textrm{cm}^{-6} \,\textrm{pc}$. While the emission measure of the Magellanic corona has not been measured, if it is similar to the MW with $0.005\le \rm EM\le0.0005$ $\textrm{cm} ^{-6}\,\textrm{pc}$ \citep{Henley2010}, it is far below the threshold for WHAM to isolate with \ha\ observations. While the observations in this paper do not trace the corona, we do find an extended, complex, multi-phase halo surrounds the LMC. 

\citet{Lehner2007} suggest the warm-hot corona gas surrounding the MCs acts as shield around stripped gas from the Clouds, preventing the MW hot corona from fully ionizing the structures and allowing them to persist longer. Our observations of the near LMC portion of LA 1.1 shows the region to be 15\%--47\% ionized (Table \ref{tab:ModelTable}). This may suggest an envelope of ionized gas surrounding LA 1.1. The presence neutral gas in this region does not directly indicate a Magellanic Corona, but additional studies of the ionizing radiation in this region may indicate if the Corona is contributing.

The total combined ionized gas mass of the central galaxies from \citetalias{Smart2019} and this study is $M_{\rm LMC+SMC,\,ionized}=\left(1.2-2.4\right)\times 10^{9}\,\msun$. This value is comparable to the mass needed in the Magellanic Corona for the LA to survive \citep{Lucchini2020}. While the models only considered the  Magellanic Corona contributing to the missing mass budget of the MC system and the ionized gas needed to shield the LA, combining both the mass from the WIM as well as the Magellanic Corona may further guide simulations tracing the history and evolution of the Clouds.

\section{Summary \label{sub:SummaryLMC}}

Using WHAM, we have mapped the extended \ha\ halo of the LMC covering 804 square degrees.
The observations cover  $+150 \leq \vlsr \leq +390$ \kms, corresponding to $-130 \leq \vlmcsr \leq +110$ \kms.
 We compare these observations to the 21-cm emission from the
HI4PI \hi\ survey and examine
the extent, morphology, velocity gradients, and mass of these two gaseous components. The main conclusions from our work are: 

\begin{enumerate}
\item \textbf{\nhi\ and \iha\ Distributions}: We see a correspondence between \ha\ emission above 0.3 R and $\log{\left(\nhi/\rm cm^{-2}\right)} \geq 20.3$ (Figure~\ref{fig:Dist}).
Outside the denser region of the LMC, continuous emission appears throughout the LMC halo above $I_{\rm H\alpha}>0.1\,\rm R$. Many of these regions
do not have neutral gas components above $\log{\left( \nhi/\rm cm^{-2}\right)} >  18.0$.
We detect a filament of ionized gas that extends from the LMC into the Leading Arm 1.1 region. 

\item \textbf{Velocity Distribution}: The ionized gas with \ha\ intensities above $I_{\rm H\alpha}>0.5\,\rm R$ 
appears to have similar kinematics to neutral gas with $\log{\left(\nhi/\rm cm^{-2}\right)} \geq 20.7$.
(Figure~\ref{fig:Vel}). In the LMC halo, the ionized gas kinematics vary more from the median \hi\ velocities. Along the LA, the ionized gas velocity
is similar to the gas velocity seen inside the LMC, suggesting the ionized gas is associated with material originating from the LMC.

\item \textbf{Ionized Gas Mass}: If we assume a distance of $D_\odot=50~\rm kpc$, we find an ionized gas mass of the central LMC
to be $M_{\rm ionized}\approx\left(6 - 18\right)\times 10^{8}\,\msun$ compared to a neutral mass
of $M_{\rm neutral}\approx\left(7.6-8.5\right) \times 10^{8}\,\msun$. The ionization fraction
ranges from 46\% to 68\%, but in all scenarios, ionized gas appears to be a major component of the LMC.
 Assumptions for $n_{e}$ and line-of-sight distances
dominate our uncertainty in the mass calculations. 

\end{enumerate}

\vspace*{1ex}

\begin{acknowledgements}
We acknowledge the support of the U.S. National Science Foundation (NSF) for WHAM development, operations, and science activities. The survey observations and work presented here were funded by NSF awards AST 1108911, 1714472/1715623, and 2009276. D.K. is supported by an NSF Astronomy and Astrophysics Postdoctoral Fellowship under award AST-2102490. Thanks to Sean Points and Frank Winkler for their help utilizing the MCELS dataset. We also thank the excellent, responsive staff at CTIO for their continued support of our remote operations.

\end{acknowledgements}

\facility{WHAM}

\bibliography{LMCPaper-aastex}
\bibliographystyle{aasjournal}

\end{document}